\documentclass[aps,prx,twocolumn,showpacs,superscriptaddress,email]{revtex4-1}
\usepackage[latin1]{inputenc}
\usepackage{amsmath} 
\usepackage{graphicx} 
\usepackage{float}
\usepackage{dcolumn}   
\usepackage{bm}        
\usepackage{amssymb}   
\usepackage[latin1]{inputenc}

\usepackage{xcolor}
\begin{document}

\title{\bfseries{Graphene, plasmons and transformation optics}}
\author{P. A. Huidobro}
\email[p.arroyo-huidobro]{@imperial.ac.uk}
\affiliation{Imperial College London, Department of Physics, The Blackett Laboratory, London SW7 2AZ, UK}
\author{M. Kraft}
\affiliation{Imperial College London, Department of Physics, The Blackett Laboratory, London SW7 2AZ, UK}
\author{R. Kun}
\affiliation{School of Precision Instrument and Opto-electronics Engineering, Tianjin University, Tianjin Tianjin 300072, P.R. China}
\author{S. A. Maier}
\affiliation{Imperial College London, Department of Physics, The Blackett Laboratory, London SW7 2AZ, UK}
\author{J. B. Pendry}
\affiliation{Imperial College London, Department of Physics, The Blackett Laboratory, London SW7 2AZ, UK}

\date{\today}

\begin{abstract}
Here we study subwavelength gratings for coupling into graphene plasmons by means of an analytical model based on transformation optics that is not limited to very shallow gratings. We consider gratings that consist of a periodic modulation of the charge density in the graphene sheet, and gratings formed by this conductivity modulation together with a dielectric grating placed in close vicinity of the graphene. Explicit expressions for the dispersion relation of the plasmon polaritons supported by the system, and reflectance and transmittance under plane wave illumination are given. We discuss the conditions for maximising the coupling between incident radiation and plasmons in the graphene, finding the optimal modulation strength for a conductivity grating. 
\end{abstract}

\maketitle

\section{introduction}

Graphene, a one atom thick layer of carbon atoms arranged in a honeycomb lattice  \cite{CastroNeto2009}, features unique optical \cite{Bonaccorso2010} and optoelectronic properties \cite{Liu2011}. In particular, it absorbs $\pi\alpha_0\sim2.3\%$ of visible light, with $\alpha_0=e^2/(\hbar c)$ being the fine structure constant. 
The conductivity of this two-dimensional (2D) material is very sensitive to external fields, such that its optoelectronic properties can be precisely tuned. A finite chemical potential $\mu\ne0$ applied to a graphene sheet, by means for instance of a gate voltage, provides a conduction band for the electrons, allowing for plasmons supported by the graphene. Remarkably, the plasmonic properties of graphene can be controlled by tuning $\mu$, and for this reason graphene has attracted a lot of attention as a highly versatile plasmonic material \cite{Vakil2011,Grigorenko2012,GarciadeAbajo2014,Low2014}.
Plasmons in graphene feature a deep subwavelength confinement together with a strong enhancement of the electromagnetic fields, as has been shown both theoretically  \cite{Shung1986,Vafek2006,Hanson2008,Jablan2009,Dubinov2011,Koppens2011,Nikitin2011} and experimentally \cite{Fei2011,Fei2012,Chen2012}. Such high field enhancement provided by graphene plasmons has been employed to increase the low optical absorption of this 2D material, which has a great potential for graphene-based optoelectronics. Different ways of coupling incident radiation into surface plasmons have been suggested: graphene sheets with modulated optical conductivity \cite{Bludov2012,Alonso-Gonzalez2014} or relief corrugations \cite{Slipchenko2013,Smirnova2015}, graphene placed on subwavelength dielectric gratings \cite{Zhan2012}, and patterned graphene structures including arrays of 1D micro-ribbons \cite{Ju2011,Nikitin2012b,Nikitin2014} and 2D micro-disks \cite{Nikitin2012,Thongrattanasiri2012a,Stauber2014a}. 

In this paper, we present an analytical formalism to study the optical response of subwavelength graphene gratings. We consider two systems: (i) a graphene conductivity grating (i.e., graphene with periodically modulated carrier density) and (ii) a graphene conductivity grating together with a dielectric grating. We make use of transformation optics \cite{Ward1996,Pendry2006,Pendry2012,Luo2014} to derive the optical spectrum of such gratings within the quasistatic approximation, study their plasmonic response, and find the optimal configurations for coupling into the surface plasmons supported by the graphene. Our method presents an important advantage to previous analytical approaches to subwavelength gratings in graphene \cite{Slipchenko2013,Zhan2012}, namely that it is not limited to very shallow grating modulations.  

\begin{figure}[b]
	\centering
	\includegraphics[width=0.99\columnwidth]{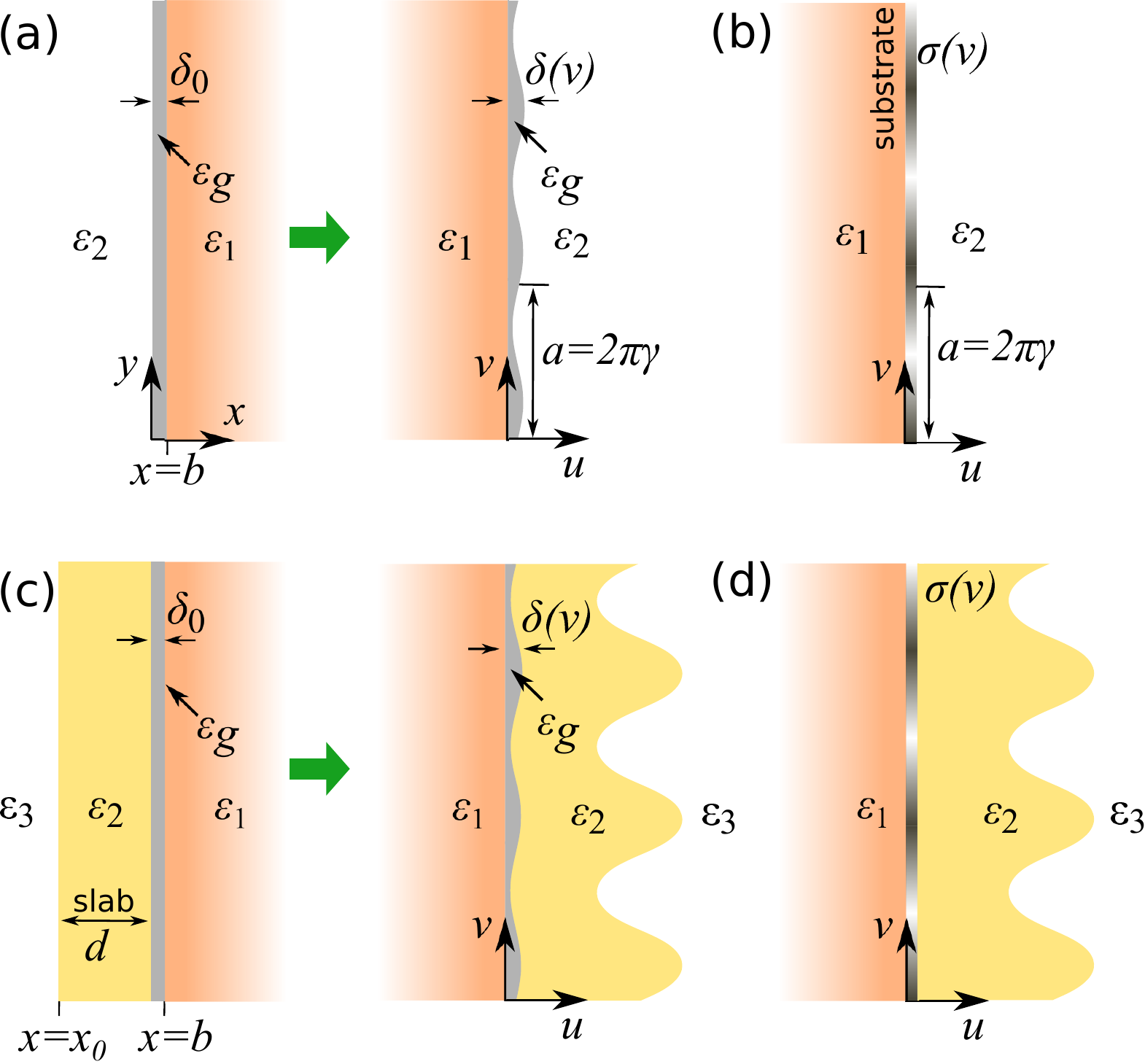}
	\caption{Effect of the conformal transformation. (a) A flat graphene sheet between two dielectric media transforms to a graphene sheet of periodically modulated thickness while it keeps the same permittivity, $\epsilon_g$. This is equivalent (b) to a periodic modulation of the sheet's conductivity, $\sigma(v)$, which can be achieved by means of a 1D modulated bias. (c) A dielectric slab on top of the graphene sheet transforms to a dielectric grating, as its surface gets modulated following the conformal map. (d) The resulting system is a graphene sheet with periodically modulated conductivity plus a dielectric grating.}
	\label{fig1}
\end{figure}


Transformation optics has successfully provided analytical solutions to various problems in plasmonics \cite{Aubry2010c,Huidobro2010,Liu2010,Pendry2013,Kraft2014,Luo2014a,Kraft2015}. Its potential resides in its ability to relate highly symmetric structures to more complex ones. In this paper we relate a simple flat graphene sheet surrounded by dielectrics to different subwavelength grating configurations of period $a$, with $a$ much smaller than the wavelength of light $a\ll \lambda$ (see Fig. 1). For this purpose, we make use of the following conformal transformation:
\begin{equation}
	w=\gamma\log\left( \frac{1}{e^z-iw_0}+iy_o\right)\, , \label{Eq.transformation}
\end{equation}
with $z=x+iy$ and $w=u+iv$. The modulation depth is determined by $w_0$, $y_0$ is given by $y_0=\frac{w_0}{e^{2b}-w_0^2}$, and the factor $\gamma$ determines the overall size of the structure. The effect of the transformation above is to map a regular Cartesian mesh into wavy contour lines that repeat themselves with period $a=2\pi\gamma$ (see Ref. \cite{Kraft2015} for further details).

\section{conductivity grating}

We start from a simple homogeneous graphene sheet placed between two dielectrics of permittivities $\epsilon_1$ and $\epsilon_2$ [see Fig. 1(a)]. The graphene can be equivalently modelled as an infinitely thin sheet with surface conductivity $\sigma_g$ or as a thin layer of thickness $\delta_0$ and with permittivity $\epsilon_g$. Let us focus on the latter situation, and assume a graphene sheet delimited by the lines at $x=b$ and $x=b-\delta_0$. Its permittivity can be related to the conductivity by,
\begin{equation}
	\epsilon_g = 1 + i\frac{\sigma_g}{\omega\epsilon_0\delta_0}\, , \label{Eq.permittivity}
\end{equation}
The effect of the transformation (\ref{Eq.transformation}) on this system is such that in the transformed space the graphene layer acquires a periodically-modulated thickness, $\delta(v)= |u(x=b)-u(x=b-\delta_0)|$, of period $2\pi\gamma$. Because the transformation relating the two frames is a conformal map, the in-plane components of the electric permittivity and magnetic permeability are conserved \cite{Pendry2002,ConfMap}, meaning that the graphene layer in the transformed space also has permittivity $\epsilon_g$ as given by Eq. \ref{Eq.permittivity}. This implies that in the transformed frame the conductivity of the graphene is periodically modulated [see Fig. 1(b)] according to, 
\begin{equation}
	\sigma(v)=\sigma_g\frac{\delta(v)}{\delta_0\gamma}\, ,
\end{equation}
where $\gamma$ is the scaling factor of the transformation. Hence, by choosing the parameters of the transformation, we can model periodically-doped graphene sheets of large modulations, controlled by $w_0$. 
Finally, it is worth noting that the conductivity modulations considered here correspond to thicknesses modulations that are much smaller than any other length scale in the system, including the plasmon wavelength.

\begin{figure}[hb]
	\centering
	\includegraphics[width=0.99\columnwidth]{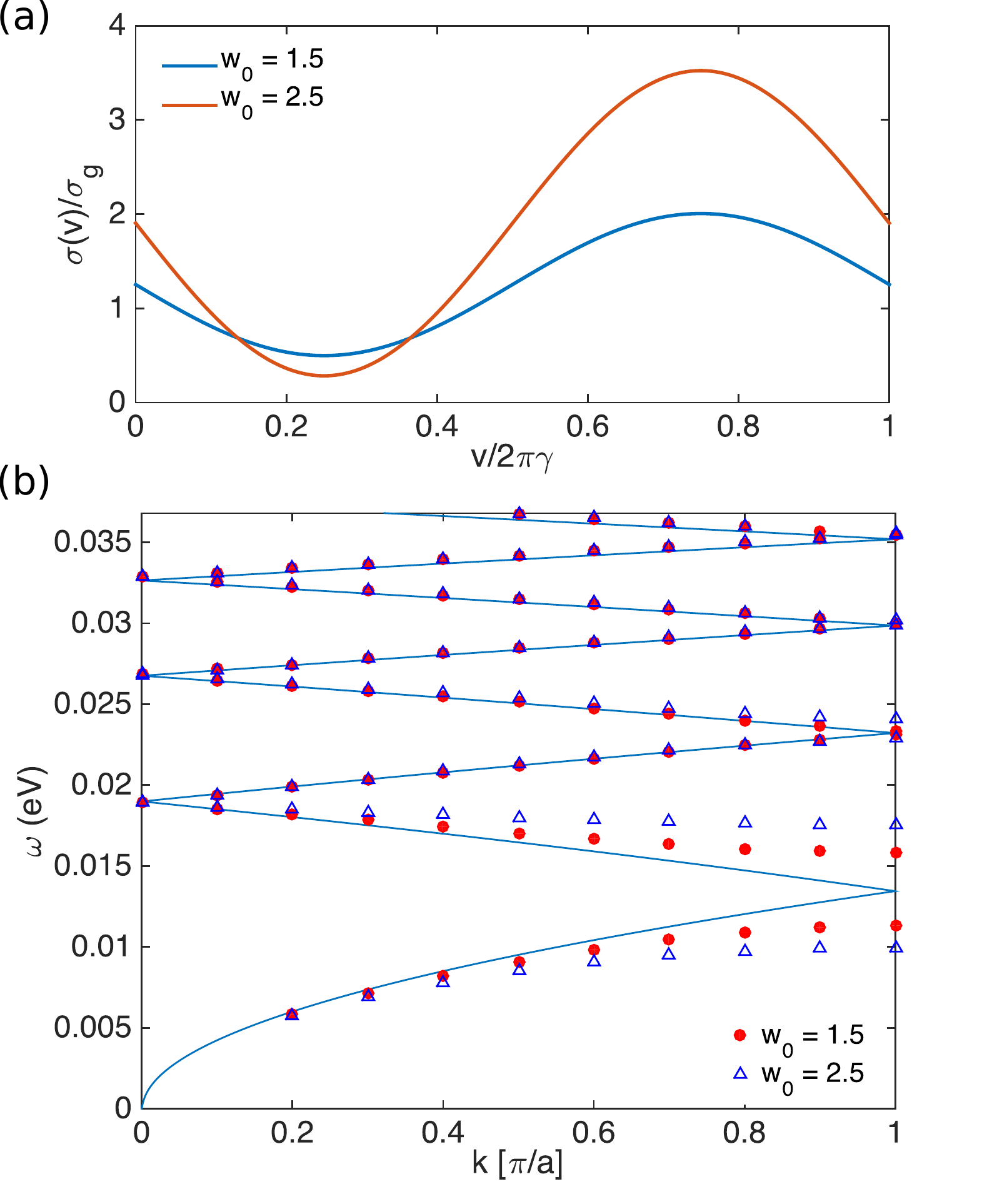}
	\caption{(a) Conductivity modulation of the graphene sheet in the transformed space for two different modulation strengths, $w_0 = 1.5$ and  $w_0=2.5$. Both profiles were derived from two lines separated by $\delta_0=2\times10^{-3}\gamma$ in the original frame, with $\gamma=4\times10^{-7}$ m. (b) Dispersion relation in the first Brillouin zone for a graphene conductivity grating at $\mu=0.1$ eV, for the conductivity profiles shown in (a). Red circles: weakly modulated conductivity grating ($w_0 = 1.5$ ). Blue triangles: strongly modulated conductivity grating ($w_0 = 2.5$).}
	\label{fig2}
\end{figure}

The periodic conductivity profile of the graphene in the transformed space is presented in Fig. 2(a) for two different modulation strengths, $w_0 = 1.5$ and $w_0 = 2.5$, which we will refer to as the weak and strong modulation cases. These modulation strengths correspond to modulation depths on the order of the period, in particular $0.75a$ and $1.6a$. Such conductivity profiles can be generated by biasing the graphene with a periodic electrostatic field. In addition, even though they were derived from the conformal transformation (1), they can be very accurately approximated to a sinusoidal shape for any value of the modulation strength. For moderate modulations, $w_0\lesssim 2$, an explicit expression for the coefficient of such sine function can be given. As we show in the SM, see Eq. S.16, we can write to zeroth order in $v$
\begin{equation}
	\sigma(v) 	\propto  \sigma_g 2i \left(d_{1}^{+}e^{b}-d_{1}^{-}e^{-b}\right)\sin(v/\gamma)   \, ,
\end{equation}
where $d_{1}^{+}=\frac{-i}{2}\frac{-1}{(w_{0}+1/y_{0})}$ and $d_{1}^{-}=\frac{-i}{2}w_{0}$ are the first order Fourier coefficients of the coordinate transformation (see SM, sections B and C). According to the above analytical expression, the conductivity modulation depends only on the geometrical parameters -- $\gamma$, $w_0$, $y_0$, $b$ -- and the conductivity of homogeneously biased graphene, $\sigma_g$, which depends on the frequency $\omega$, chemical potential $\mu$ and temperature $T$ -- we use $T=300$ K throughout the paper --.  In our simulations we use the conductivity of graphene as given in the random phase approximation, with intraband and interband contributions,  $\sigma_g=\sigma_g^{\text{intra}}+\sigma_g^{\text{inter}}$, which read as follows, \cite{wunsch2006} 
\begin{eqnarray}
	\sigma_g^{\text{intra}} &=&\frac{2ie^2t}{\hbar\pi\left[\Omega +i\gamma\right]}\,\text{ln}\left[ 2\,\text{cosh}\left( \frac{1}{2t} \right)\right], \label{grapheneConductivity1} \\
	\sigma_g^{\text{inter}}&=&\frac{e^2}{4\hbar}  \label{grapheneConductivity2} \\
	 \left[ \frac{1}{2} \right. &+& \left.\frac{1}{\pi} \text{arctan} \left( \frac{\Omega-2}{2t} \right) - \frac{i}{2\pi} \text{ln} \frac{(\Omega+2)^2}{(\Omega-2)^2 +(2t)^2} \right].  \nonumber
\end{eqnarray}
Here we have introduced a normalized frequency, $\Omega=\hbar\omega/\mu$, and temperature $t=k_BT/\mu$. The damping term, $\gamma=\hbar/(\mu\tau)$, is given by the carriers' scattering time, $\tau=m\mu/v_F^2$ ($m$ is the mobility and $v_F$ the Fermi velocity). We refer the reader to the Supplementary Material (SM) for more details. 

The periodic modulation of the conductivity acts as a grating that supplies the momentum mismatch for radiation to couple into surface plasmons. Since the momentum of plasmons in graphene is much larger than that of free space radiation, the grating period needs to be much smaller than the incident wavelength ($2\pi\gamma\ll \lambda$). Note here that similar systems, where the coupling to graphene plasmons is provided by periodically patterning the graphene has received large attention \cite{Ju2011,Nikitin2012b,Nikitin2014,Nikitin2012,Thongrattanasiri2012a,Stauber2014a}. In particular, 1D arrays of micro ribbons, present similar plasmon resonances in the same frequency range for a given periodicity.

\subsection{Plasmon modes and resonance condition}

In the electrostatic limit the spectral properties of a plasmonic structure depend only on its geometry. As it was recently shown, transformation optics is specifically suited to classify plasmonic resonances in terms of geometrical symmetries, since it is able to reveal `hidden' symmetries present in plasmonic structures \cite{Kraft2014}. The reason for this is that conformal transformations conserve not only the in-plane $\epsilon$ and $\mu$ but also the electrostatic potential. This implies that the plasmon modes and resonance condition in the transformed space are directly given by those in the simple original frame. 

The plasmon resonance condition of a translationally invariant graphene sheet between two dielectrics is well known and, under the quasistatic approximation, reads as, 
\begin{equation}
	\epsilon_1+\epsilon_2+ 2i\alpha\frac{|k|}{k_0}=0\, , \label{Eq.dispersion}
\end{equation}
where $k$ is the mode's parallel momentum, $k_0=\omega/c$ and we have introduced a normalized conductivity for graphene $\alpha=2\pi\sigma_g/c$. On the other hand, by the argument above, all the graphene conductivity gratings related to this system by  transformation (\ref{Eq.transformation}) also satisfy the same plasmon resonance condition. This means that the spectrum of a whole class of conductivity gratings, characterized by the modulation strength, $w_0$, and period, $2\pi\gamma$, can be derived \cite{Kraft2015} from the simple plasmon condition of a graphene sheet given in Eq. \ref{Eq.dispersion}. 

Figure 2(b) presents the dispersion relation in the first Brillouin zone for homogeneously doped graphene on a substrate, as obtained from the analytical expression Eq. \ref{Eq.dispersion} (solid line). The substrate has a permittivity of $\epsilon=3$, similar to typical values for polymers in the THz regime. Here and throughout this work we take the graphene's conductivity as given by the random phase approximation and with a scattering time for the electrons of $\tau=10$ ps
(see SM section A for comments on the losses). The graphene sheet is subject to a chemical potential of  $\mu=0.1$ eV, i.e. in the THz regime and far below phonon losses ($\mu  \gtrsim 0.2 $ eV ) \cite{GarciadeAbajo2014}. Note that the operating frequency regime of the systems under study can be tuned by changing $\mu$ or by rescaling the structure.  

The plot also shows the band structures of two different conductivity gratings belonging to the same class and for the same parameters used in panel (a). These were obtained from electrodynamics simulations (Comsol Multiphysics) and are plotted with dots. Because the system in the transformed space is periodic along the $v$ axis, we need to impose the same periodicity in the $(x,y) $ frame, where the dispersion relation is continuous at the zone edge. However, solutions at a finite wave vector have an unphysical discontinuity of the phase across the branch cuts of Eq. \ref{Eq.transformation}. For this reason, the equivalence between the homogeneously doped graphene and the modulated graphene strictly holds only at the zone centre, and the conductivity grating case features a band gap opening at the zone edge. The strongest effect is for the first order mode and for the strongest modulation. This is because the conductivity profiles derived from the conformal transformation and shown in Fig. 2(a) can be well approximated by a sinusoidal function, such that these gratings mostly provide coupling to the first order Fourier component. In fact, all the higher order Fourier coefficients are several orders of magnitude smaller than the first order one (see Fig. S2 in SM). In addition, the quasistatic approximation disregards magnetic effects, which account for small discrepancies between the band structures in the original and transformed frames. In particular, very small band gaps (not shown here) open at $k=0$ (see Ref. \cite{Kraft2015} for further comments). In any case the magnetic gaps are much smaller than the broadening of the bands due to losses and therefore have insignificant effects.

\subsection{Optical response of the grating under plane wave illumination}
We now use transformation optics to calculate quantities that are attainable in experiments. In particular, we derive analytical expressions for the transmission and reflection spectra of graphene conductivity gratings under plane wave illumination at normal incidence. Only the main steps of the calculation are presented here, while a more detailed derivation can be found in the SM section E.   

We start by considering a plane wave incident from the right in the grating frame, $(u,v)$, and, following Ref. \cite{Kraft2015}, we approximate it in the vicinity of the graphene grating, 
\begin{eqnarray} 
	\mathbf{H}^{sou} &=& -\frac{\omega\epsilon_2\epsilon_0}{k}E^{sou}e^{-iku}\mathbf{z}  \\ 
	&=& -\frac{\omega\epsilon_2\epsilon_0}{k}E^{sou}(1-iku)\mathbf{z}.
\end{eqnarray}
with $k=\sqrt{\epsilon_2}k_0$. Similar to the electrostatic potential, the $z$-component of the magnetic field is conserved under a conformal transformation \cite{Ward1996,Leonhardt2008a}. This property, together with the Fourier expansion of the coordinate transformation (1), allows us to write the incident potential in the original space as, 
\begin{eqnarray}
	&&  \phi^{sou} = E^{sou}\gamma  \nonumber \\ 
	&& \times \left[ \underset{g\neq0} {\sum_{g=-\infty}^\infty}   i\text{sign}(g)\left(  d^+_g e^{|g|x} -   d^-_g e^{-|g|x}  \right) e^{igy} + y   \right] \label{Eq.sourcepotential}
\end{eqnarray}
for $\log(w_0)<x<\log(w_0+1/y_0)$. The expansion coefficients $d_g^{\pm}$ derive from the coordinate transformation and are given in the SM. It is clear from the expression above that a plane wave with only one Fourier component in the grating frame, is transformed to a wave that contains all higher order modes in the frame where the conductivity of graphene is homogeneous. In other words, the conductivity grating provides coupling between plane waves and modes bound to the graphene. 
The total electrostatic potential at both sides of the graphene can then be expressed as, 
\begin{eqnarray}
	\phi_L &=& \phi_L^{sou} +\phi_L^{near} + \phi_L^{rad} \nonumber \\
	\phi_R &=& \phi_R^{near} + \phi_R^{rad} 
\end{eqnarray}
where the subindices $L$ and $R$ stand for the half-spaces  left and right  to the graphene sheet. The above expression includes the following terms (full expressions are given in the SM): (i) a near field part, $\phi_{L,R}^{near}$ that arises from the contribution of the scattered evanescent electrostatic modes from the graphene; and (ii) a radiative part, $\phi_{L,R}^{rad}$, that originates from the fact that the conductivity grating induces a continuous current in the grating, which in turn radiates outgoing plane waves with amplitudes $E^{ref}$ and $E^{tra}$:
\begin{eqnarray} 
	\mathbf{H}^{ref} &=& \frac{\omega\sqrt{\epsilon_2}\epsilon_0}{k_0}E^{ref}e^{i\sqrt{\epsilon_2}k_0u}\mathbf{z}  \\
	\mathbf{H}^{tra} &=& -\frac{\omega\sqrt{\epsilon_1}\epsilon_0}{k_0}E^{tra}e^{-i\sqrt{\epsilon_1}k_0u}\mathbf{z} 
\end{eqnarray}
It is important to note that the radiative part is not present in a purely electrostatic calculation and incorporates the radiative reaction of the conductivity grating into our theoretical formalism.

The amplitude and expansion coefficients are then obtained by imposing the boundary conditions for the EM fields at $x=b$. These are the continuity of the tangential electric field, $E_{y,R} - E_{y,L} = 0$, and the discontinuity of the normal displacement field, $ D_{x,R} - D_{x,L} = \Sigma $. The surface charge density on the graphene, $\Sigma$, is obtained from the continuity equation, $-i\omega\Sigma + \nabla\mathbf{j}=0$, where the surface current density along the graphene is $\mathbf{j}=\sigma_gE_y\mathbf{y}$. 

\begin{figure}[bh]
	\centering
	\includegraphics[width=0.99\columnwidth]{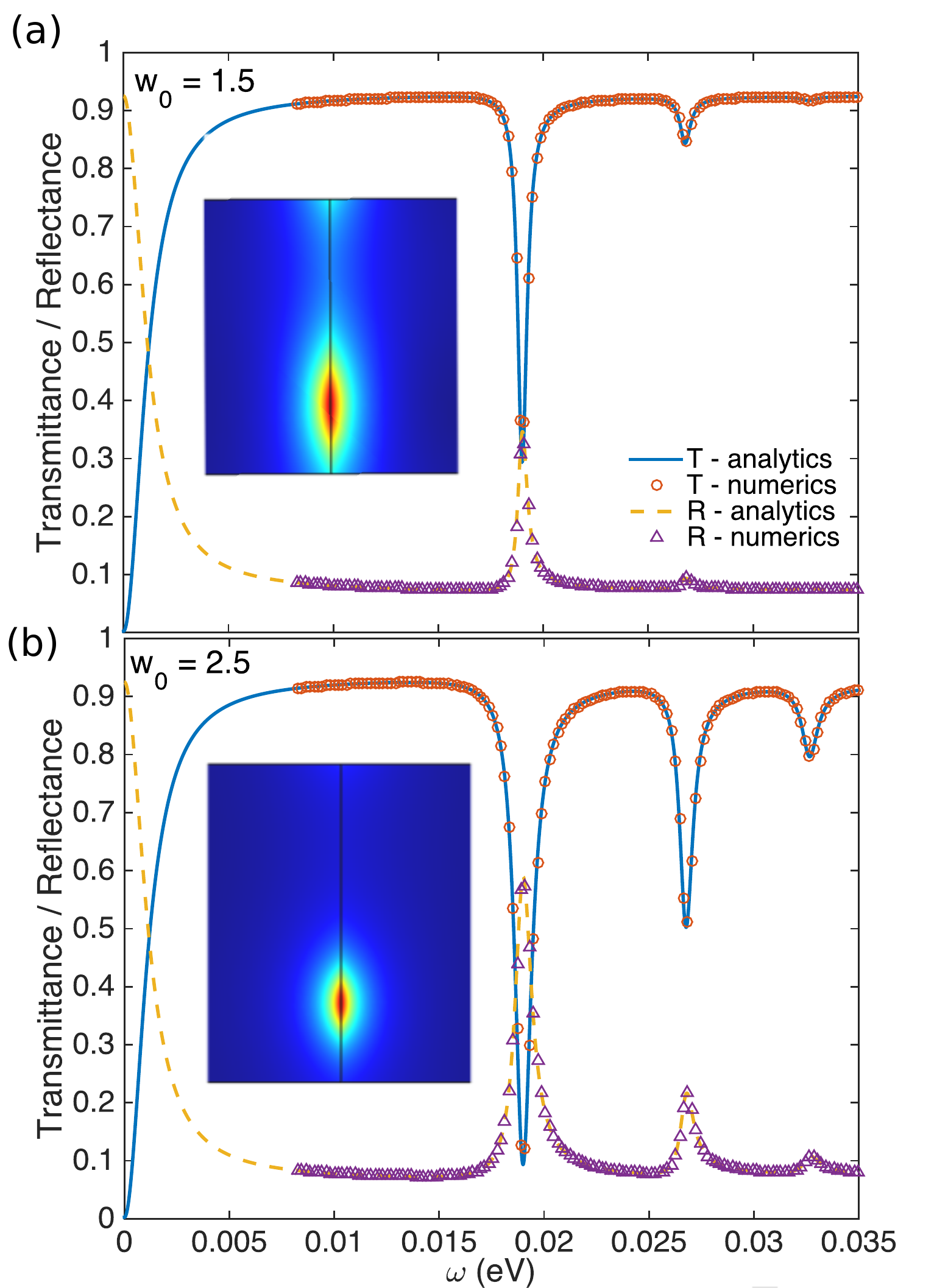}
	\caption{Graphene conductivity grating: The conductivity of the graphene sheet is periodically modulated following the profiles shown in Fig. 2(a). Optical response at normal incidence: Transmittance and reflectance spectra for two different modulation strengths, (a) $w_0 = 1.5$ and (b) $w_0=2.5$. The inset panels show the respective norm of the electric field at the first order plasmon resonance. The maximum value of $|\mathbf{E}|/|\mathbf{E_0}|$ is 30 in (a) and 55 in (b). The graphene is placed on top of a substrate ($\epsilon_1=3$, $\epsilon_2=1$) and illuminated from the free space side. }
	\label{fig3}
\end{figure}

In order to determine all coefficients unambiguously, we introduce an additional radiation boundary condition \cite{Kraft2015}. This is related to the radiative reaction of the graphene conductivity grating and is imposed in the grating frame. According to Maxwell's equation $\mathbf{\nabla}\times\mathbf{H}=\mathbf{j}$, 
the tangential component of the magnetic field is discontinuous across the thin current sheet,  
\begin{equation} 
 	H_z^{tra} - H_z^{sou} - H_z ^{ref} = j^c \,. 
\end{equation}
Here, $j^c$ is the total conduction current density of the graphene layer in the transformed frame. It can be obtained by writing the current carried by the graphene sheet in the original space, 
\begin{equation}
 J_y^c = \sigma_g\delta(x-b)E_y \, ,
 \end{equation} 
transforming it into the grating frame, $J_v^c$, integrating it in the whole unit cell, and normalising to the unit cell area. This procedure yields the total surface current density,
\begin{eqnarray}
j^c &=& \sigma_gE^{tra} N \, , \\ 
	 N &= &1+ \underset{g\neq0} {\sum_{g=-\infty}^\infty}  |g|^2 \left[  i\text{sign}(g) ( h_g^+ d_{2,-g}^+ e^{2|g|b}  - h_g^- d_{2,-g}^- e^{-2|g|b}     \right. \nonumber \\
	&& \left.  - h_g^+ d_{2,-g}^- + h_g^- d_{2,-g}^+)  - (h_g^+ + h_g^- e^{-2|g|b} e^{sca}_{2,-g} )  \right]
\end{eqnarray}
Note that $N$ contains a sum over all higher order Fourier components, meaning that they all contribute to the current in the grating. 

From the boundary conditions and the expression for the current we determine the reflection and transmission coefficients to be
\begin{eqnarray} 
	r &=&  \frac{\sqrt{\epsilon_1}-\sqrt{\epsilon_2}-2\alpha N}{\sqrt{\epsilon_1} +\sqrt{\epsilon_2} + 2\alpha N} \, , \label{Eq.r}\\ 
	t &=&  \frac{2\sqrt{\epsilon_1}}{\sqrt{\epsilon_2} +\sqrt{\epsilon_1} + 2\alpha N}\label{Eq.t} \,.
\end{eqnarray}
Finally, reflectance and transmittance are given by
\begin{eqnarray}
	R & = & |r|^2 \, , \\
	T & = & \frac{\sqrt{\epsilon_1}}{\sqrt{\epsilon_2}}|t|^2  \, .
\end{eqnarray}

The reflection and transmission properties of the two conductivity gratings considered earlier are shown in \mbox{Fig. 3}. The plots show the reflectance and transmittance for two graphene gratings with weak (a) and strong (b) modulations illuminated from the right. In both cases the agreement between the analytical results obtained with equations \ref{Eq.r} and \ref{Eq.t} (plotted with lines) and the numerical results (represented with dots) is nearly perfect. Our analytical model predicts very accurately not only the position of the resonances, as was clear from Fig. 2, but the full EM response at normal incidence. This confirms that, up to the quasistatic limit, it is only the properties of the untransformed graphene sheet that determines the response of these kind of grating.  On the other hand, from the field intensity plots given as inset panels, we observe that these graphene conductivity gratings provide a very large field enhancement that can be used to enhance non-linear effects. In particular, for the parameters used in Fig. 3 we find that the maximum value of $|\mathbf{E}|/|\mathbf{E_0}|$ is 55 for the case with $w_0=2.5$ and 30 for $w_0=1.5$. Note that the field enhancement is reduced when lower scattering times are considered in the simulations. In particular, for $\tau=1$ ps, the maximum field enhancement values for the strong and weak modulations are 35 and 17, respectively, while for $\tau=0.1$ ps, we obtain 7 and 3.

\subsection{Coupling optimization and Impedance matching}

\begin{figure}[t]
	\centering
	\includegraphics[width=0.99\columnwidth]{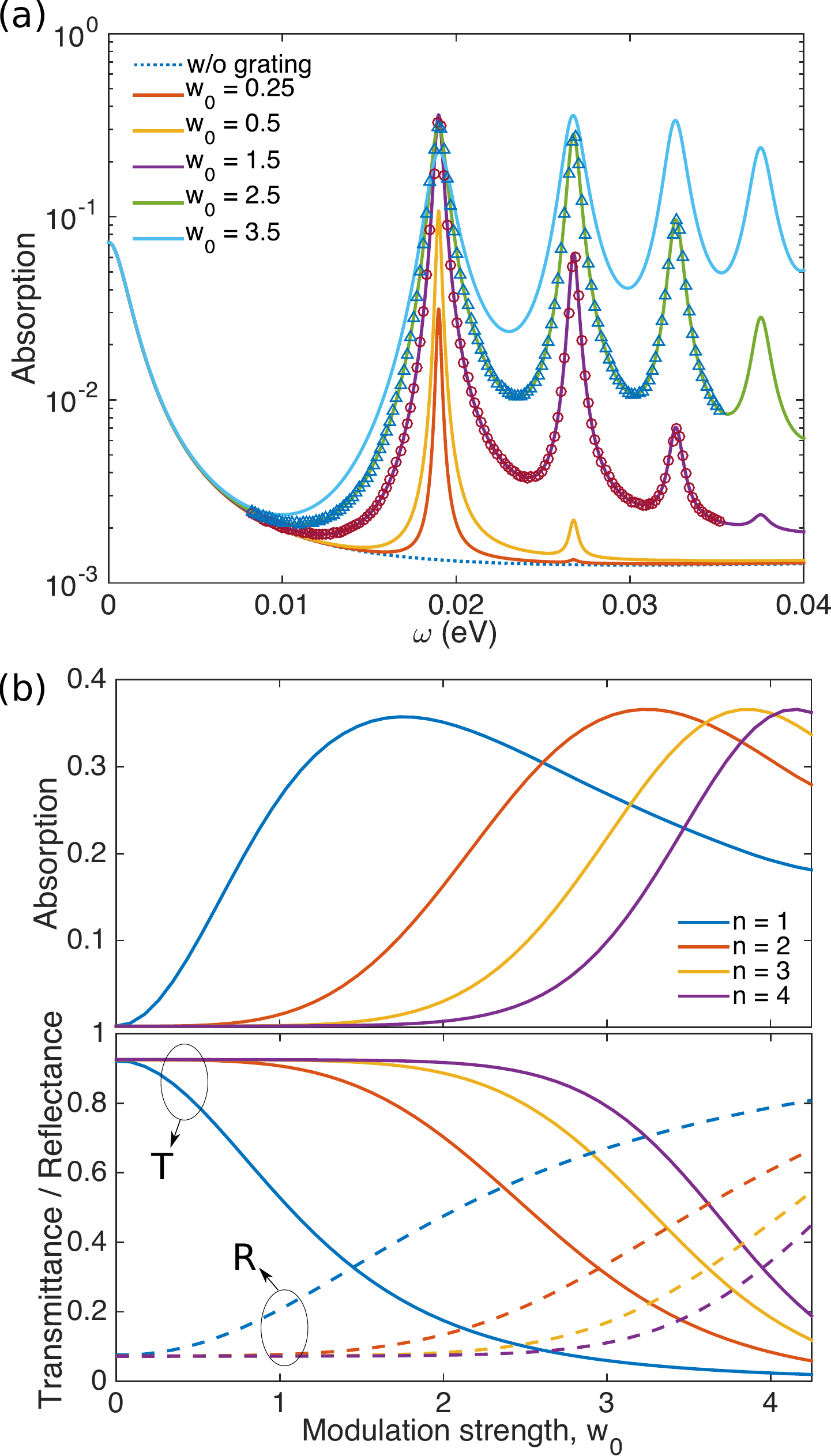}
	\caption{Optimal modulation strength for the graphene conductivity grating. (a) Absorption spectrum at normal incidence for different grating modulation strengths, $w_0$. The dots represent simulation results for $w_0 = 1.5$ and $w_0=2.5$, the two cases shown in Fig. 3. (b) Top panel: Maximum of the absorption peak as a function of $w_0$ for the different order modes. Bottom panel: minimum of the transmittance peak (solid lines) and maximum of the reflectance peak (dashed lines). The colour legend in the top panel applies to both panels.  }
	\label{fig4}
\end{figure}

We now take advantage of our analytical results to find the optimal parameters for radiation coupling into plasmon modes. 

In Fig. 4 (a) we show the absorption spectrum of a whole class of conductivity gratings with different modulation strengths, that is, different values of the $w_0$ parameter. Since all the gratings derive from the same original physical system, they all feature the same plasmon resonance frequencies. As $w_0$ increases, the coupling to the incident wave increases and the absorption spectra display higher order modes: while for the weakest modulation considered, $w_0=0.25$, only the dipole mode is visible in the spectrum, the first four modes are excited for modulation strengths larger than $w_0=1.5$. Since the coupling is enhanced as $w_0$ grows, the height of the peaks increases. However, by increasing $w_0$ also the radiation losses increase, resulting in a broadening of the peaks. For the first order mode, such an increase in radiation damping results in lower absorption maxima for the highest values of $w_0$. This is clearly shown in Fig. 4 (b), where the absorption maxima of the different modes are plotted as a function of modulation strength, and implies that there is an optimal value of the conductivity modulation strength, $w_0\approx 1.75$ for this set of parameters, that yields the highest coupling to the plasmon dipole mode. This is one of the main results of this paper. 

As in electrical circuit theory, when the impedance of the generator (in our case the radiative resistance) and that of the load are matched, energy dissipated is a maximum and is shared equally between the generator and the load. Hence in figure 4 (b) we note that at optimum the absorption is maximised to nearly $1/(1+\sqrt{3})\approx0.366$ indicating that almost perfect matching is achieved.

\section{dielectric and conductivity grating}
We now consider the second structure shown in Fig. 1: a graphene sheet with periodically modulated conductivity together with a dielectric grating of the same periodicity, as sketched in panel (d). The periodically biased graphene and dielectric grating system is placed between two dielectrics, $\epsilon_1$ and $\epsilon_3$, which we will take to be free space for simplicity. The dielectric grating consists of a thin slab on top of the graphene, with different permittivity, $\epsilon_2=\epsilon_d$, and with a periodic relief modulation of the surface. The structure under consideration results from applying transformation (\ref{Eq.transformation})
to a system composed of a flat dielectric slab of permittivity $\epsilon_2$ and thickness $d$ on top of a graphene sheet of homogeneous conductivity $\sigma_g$, as depicted in the left panel of Fig 1 (c). By means of the transformation, both the conductivity of the graphene and the surface of the dielectric slab acquire a periodic modulation of the same pitch ($a=2\pi\gamma$) and corresponding intensity (given by $w_0$). Therefore, by changing the parameters $\gamma$ and $w_0$, we can derive a whole class of graphene-plasmonic structures where coupling to external light is provided by a conductivity grating together with a dielectric grating. The fact that both gratings have the same periodicity is an advantage from the point of view of the experimental realisation of this system: by applying an electrostatic field to the dielectric slab that acts as a dielectric grating, a modulated doping can be induced in the graphene with the same periodicity.

\begin{figure}[htb]
	\centering
	\includegraphics[width=0.9\columnwidth]{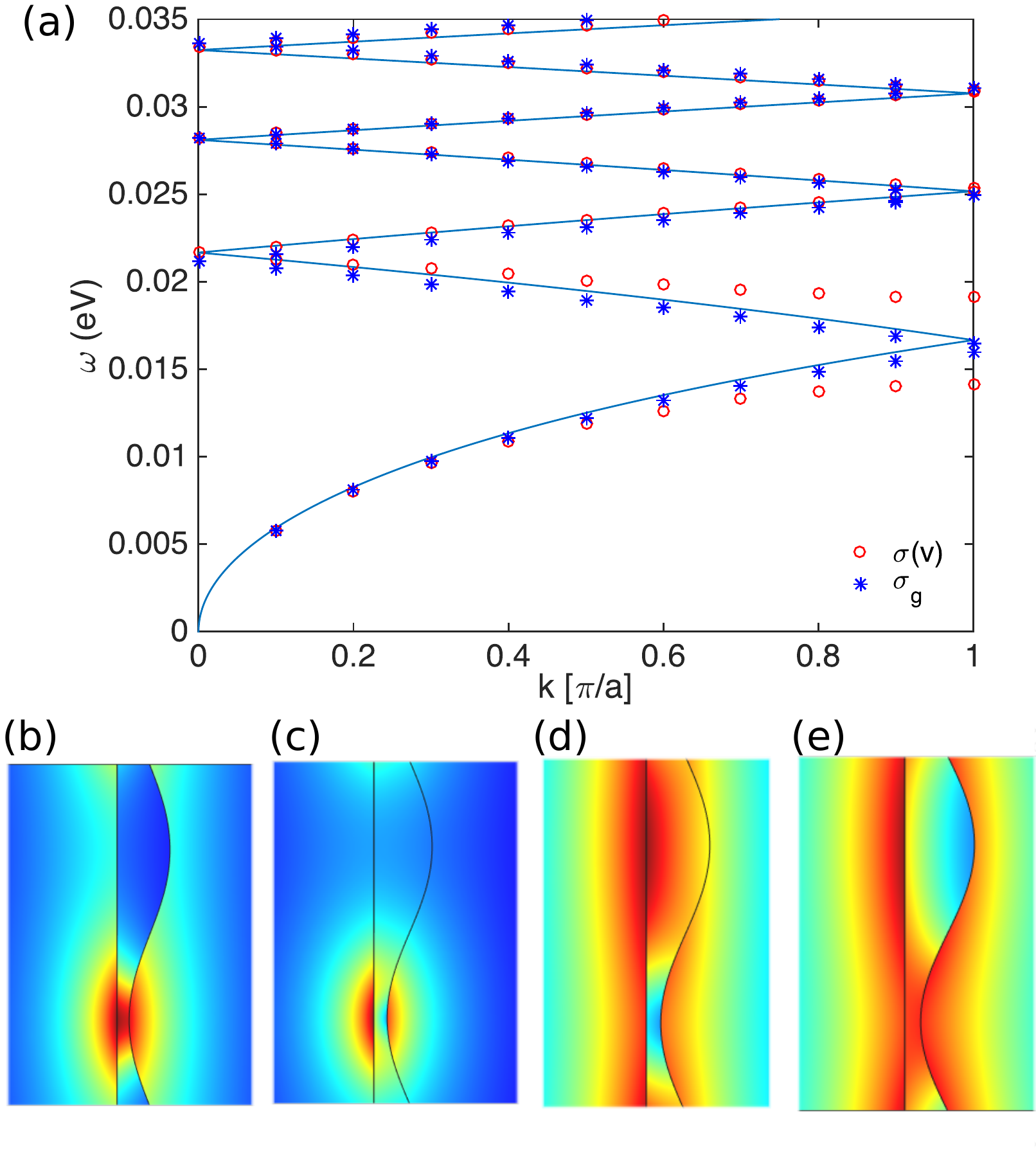}
	\caption{Subwavelength dielectric gratings for the excitation of graphene plasmons. The dielectric grating (permittivity $\epsilon_2=3$, period $2\pi\gamma$, and thickness in the original frame $d=0.5\gamma$, $\gamma=4\times10^{-7}$) is placed in free space ( $\epsilon_1=1 $, $\epsilon_3=1 $). (a) Band structure of the gratings: the red circles (blue stars) represent simulation results for the same dielectric grating with (without) a conductivity grating of the same periodicity. The modulation strength is in both cases $w_0=1.5$. The solid line plots the analytical plasmon dispersion given by Eq. \ref{Eq.dispersion2}. (b,c) Norm of the electric field for the lower energy (b) and upper energy (c) first order modes at the zone edge for the case of dielectric and conductivity gratings. (d,e) Same as (b,c) but for the case of dielectric grating and homogeneously doped graphene.}
	\label{fig6}
\end{figure}

\subsection{Plasmon modes and resonance condition}

A whole class of graphene conductivity and dielectric grating derives its optical properties from a graphene sheet of conductivity $\sigma_g$ (placed at $x=b$) and a dielectric slab of permittivity $\epsilon_d$ and thickness $d$ (placed at $x=x_0$). Therefore, the plasmon resonance condition in the slab frame, which within the quasistatic limit reads as (see SM section F),
\begin{equation}
	e^{2|k|d}=\left(\frac{\epsilon_d-1}{\epsilon_d+1}\right)\frac{(\epsilon_d-1)k_0-2i\alpha|k|}{(\epsilon_d+1)k_0+2i\alpha|k|} \label{Eq.dispersion2}
\end{equation}
also determines the dispersion relation of the plasmon modes in the grating frame.

Figure 5 presents the dispersion relation for an instance of the class of conductivity and dielectric gratings derived from the flat system (with $d = 0.5$ and $x_0 = 1$). The parameters are the same as those considered in Section II: $\mu=0.1$ eV, $\gamma=4\times10^{-7}$, $\epsilon_d=3$. The solid line represents the analytical prediction given by Eq. \ref{Eq.dispersion2}, and is to be compared with the numerical results for a grating with $w_0=1.5$ (red circles) obtained from full electrodynamics simulations. As in the conductivity grating case discussed above, the agreement between the numerical and analytical results is nearly perfect near the zone centre, while simulations reveal a band gap opening at the zone edge. Similarly to the previous case, this band gap is largest for the first order mode, but, differently, it is still appreciable for higher order modes. The reason for this is that the dielectric relief grating cannot be as accurately approximated to a $\sin(v)$ function as the conductivity grating, although it is still a good approximation (see Fig 1 in the Supplementary material). This difference stems from the position of the different starting lines in the slab frame: while the conductivity grating originates from a line that is very close to $x=b$ (the un-transformed line), the dielectric grating is obtained from a line at a distance $d$, which transforms to a line that is close to the branch points, where distortion is larger and the transformation cannot be as accurately described with a single Fourier component. 
   
In addition, in Fig. 5 we show for comparison the numerical results for the band structure of a dielectric grating on top of a homogeneously doped graphene (blue stars). In other words, we consider the same dielectric slab with a periodically corrugated surface, but this time we do not include the conductivity grating. Two remarks can be extracted from this comparison. First, the resonance condition given by the analytical expression \ref{Eq.dispersion2} fails to predict the spectral position of the modes in the zone centre as accurately as in the previous case. Second, although this case also presents a band gap at the zone edge, it is much smaller than in the case where a conductivity grating is also implemented.

This difference in the band gap opening can be understood by looking at the field profiles of the first order modes at the zone edge for the two cases [see Fig. 5 (b-e)]. When both a conductivity and dielectric grating are implemented (b,c), the modes are tightly localised at the conductivity minimum, which coincides with the thinner part of the dielectric grating. While for one of the modes the field resides mainly in the dielectric grating, for the other it peaks in the free space region, such that frequency of the first is reduced and the frequency of the second is increased, and a band gap opens. On the other hand, in the presence of only the dielectric grating, i.e., without the conductivity grating, the fields are not so tightly localised at the thin part of the grating (d,e). In this case, since the graphene is homogeneously doped, the field is more uniformly distributed along the unit cell, resulting in a much smaller band gap. It is worth noting here that periodically modulating the doping of the graphene has a strong effect on field concentration at the zone edge modes. 

\subsection{Optical response of the grating under plane wave illumination}
We now derive analytical expressions for the transmission and reflection coefficients of graphene, subject to the conductivity and dielectric gratings. We consider a plane wave impinging from the right of the structure shown in Fig. 1(d) and follow a procedure similar to that described in Section II.B. Since we are making use of the same transformation, the expansion of the incident electrostatic potential, $\phi^{sou}(x,y)$, remains the same (Eq. \ref{Eq.sourcepotential}). Based on this, we write the fields in the different regions of space as,
 \begin{eqnarray}
	\phi_L &=& \phi_L^{sou} +\phi_L^{near} + \phi_L^{rad} \nonumber \\
	\phi_M &=& \phi_M^{near} \nonumber \\
	\phi_R &=& \phi_R^{near} + \phi_R^{rad} 
\end{eqnarray} 
While the potential at the left and right free-space regions have the same expressions as before, $\phi_M$ stands for the potential within the dielectric slab and reads as, 
 \begin{eqnarray}
\phi_M  &=&  E_0^v\gamma y   \nonumber\\
	 &+& \underset{g\neq0} {\sum_{g=-\infty}^\infty}   \left(  c^+_g e^{|g|x} +  c^-_g e^{-|g|x}\right) e^{igy}  
\end{eqnarray} 
Next, we apply the boundary conditions at the two interfaces: the continuity of the tangential component of the electric field and the normal component of the displacement field at $x=x_0$; and the continuity of the tangential component of the electric field and discontinuity of the normal component of the displacement field at $x=b$.  
\begin{figure}[!b]
	\centering
	\includegraphics[width=0.9\columnwidth]{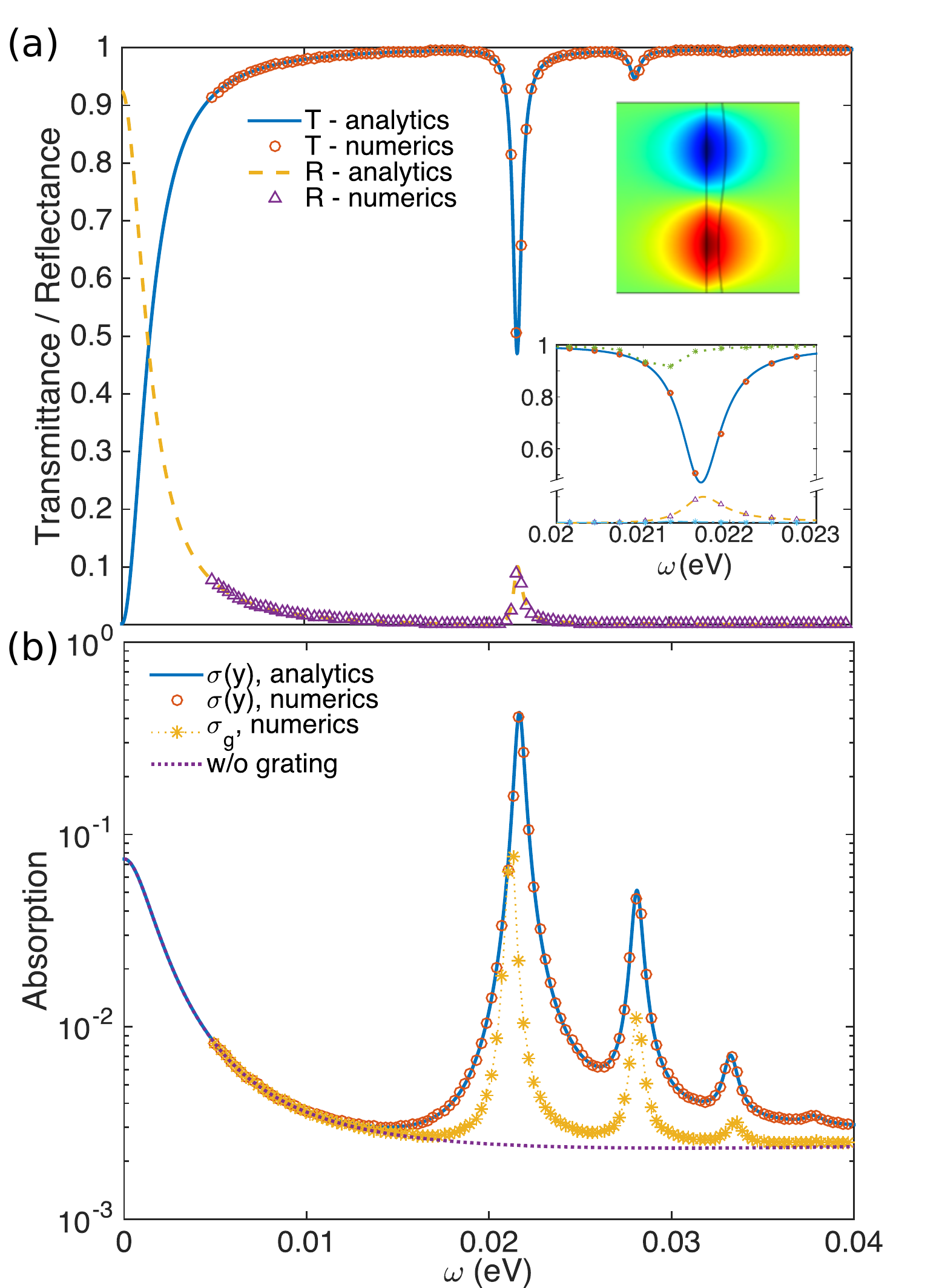}
	\caption{Optical response of the dielectric and conductivity grating system: the surface of the dielectric slab and the conductivity of the graphene sheet are modulated with the same periodicity as in Fig. 5 and for $w_0=1.5$ (a) Reflectance and transmittance at normal incidence: from the analytical expressions (solid and dashed lines) and from simulations (circles and triangles). The top inset panel shows the field profile at the dipole mode. The lower inset panel is a zoom of the first order resonance showing also for comparison the spectrum of a system with dielectric grating only, i.e., without the conductivity grating. In this case the stars represent simulation results and the dotted lines are a guide-to-the-eye. (b) Absorption spectrum for the same cases as in panel (a). The case with no coupling to plasmons is shown as a dotted line for reference. }
	\label{fig6}
\end{figure}

As a last step, we implement the additional radiation boundary condition. In this case, we consider the graphene plus dielectric system as a thin current layer, radiating outgoing plane waves, such that the magnetic field in the left and right hand sides of the structure is discontinuous across this radiating layer, $H_z^{tra} - H_z^{sou} - H_z ^{ref} = j$. Two contributions to the current need to be taken into account here: one from the conduction along the graphene sheet ($j^c$ which was also present in the previous case), and one from the displacement current within the dielectric area ($j^D$), 
\begin{equation} 
 	j = j^c +j^D.
\end{equation}
By applying the same procedure to calculate the conduction current as in the previous case we arrive at  
\begin{equation} 
 	j^c = \sigma_g(E^{sou}+E^{ref})  N^c \, ,
\end{equation}
where $N^c$ reads as
\begin{eqnarray}
	 N^c &= &1+ \underset{g\neq0} {\sum_{g=-\infty}^\infty}  |g|^2 \left( h_g^+ c_{2,-g}^+ e^{2|g|b}  + h_g^- c_{2,-g}^- e^{-2|g|b}     \right. \nonumber \\
	&& \left.  + h_g^+ c_{2,-g}^- + h_g^- c_{2,-g}^+   \right)
\end{eqnarray}
On the other hand, starting with the displacement current in the slab frame, 
\begin{equation}
 \mathbf{J^D} = i\omega(\epsilon_d-1)\epsilon_0\mathbf{E} \,,
 \end{equation} 
and following the procedure described for the conduction current, the total displacement current in the grating frame can be shown to be given by
\begin{equation} 
 	j^D = i\omega(\epsilon_d-1)\epsilon_0 \gamma (E^{sou}+E^{ref}) N^D \, . 
\end{equation}  
In this case, $N^D$ is
\begin{eqnarray} 
	 N^D &= & -d+ \underset{g\neq0} {\sum_{g=-\infty}^\infty}  |g| \left[ h_g^+ c_{2,-g}^+ (e^{2|g|b} -e^{2|g|x_0} )   \right. \nonumber \\
	&& \left.  - h_g^- c_{2,-g}^- (e^{-2|g|b} -e^{-2|g|x_0} )     \right]
\end{eqnarray}

Making use of the boundary conditions in the slab frame together with the radiation boundary condition we arrive at,
\begin{eqnarray} 
	r &=&  - \frac{ 2\alpha N^c + ik_0(\epsilon_d -1) \gamma N^D }{2 + 2\alpha N^c + ik_0(\epsilon_d -1) \gamma N^D  } \, , \label{Eq.r2}\\ 
	t &=&  \frac{2}{2+2\alpha N^c + ik_0(\epsilon_d -1) \gamma N^D }\label{Eq.t2} \,.
\end{eqnarray}
Then, reflectance and transmittance are given by $R=|r|^2$ and $T=|t|^2$. Finally, note here that our analytical model assumes that the current sheet radiates symmetrically on both sides, which will not be the case for very strongly modulated gratings or higher frequencies.

In Fig. 6 we present the optical spectrum at normal incidence for a dielectric and conductivity grating with the same periodicity and chemical potential as in Fig. 5 and for modulation strength $w_0=1.5$. The solid lines in panel (a), which depict reflectance and transmittance obtained from Eqs. \ref{Eq.r2} and \ref{Eq.t2}, show a very good agreement with the simulation results (dots), as well as those in panel (b), which represent absorption. On the other hand, we also show numerical results for the case when only a dielectric grating is considered, i.e., without modulation of the conductivity. It is clear from these results that the optimal way to couple into plasmons is to have both a dielectric and a conductivity grating, which greatly enhances absorption in the graphene and suppresses transmission through the system, as opposed to the less efficient coupling provided by only modulating the dielectric interface.

Our analytical model allows us to study the behaviour of the coupling with the modulation strength. In contrast to the case considered in the previous section, where we showed that coupling into graphene plasmons by a conductivity grating can be maximised for a range of modulation strengths, we see in Fig. 7 that for the case where a dielectric grating is included absorption increases monotonically with $w_0$. That is, the deeper the modulation, the more efficient the coupling into plasmons. For the largest modulation strengths considered absorption reaches 0.5, a signature of perfect impedance matching. The fact that there is no optimal modulation strength for more efficient coupling into plasmons is consistent with the previous analysis in Ref. \cite{Slipchenko2013}.

\begin{figure}[!ht]
	\centering
	\includegraphics[width=0.9\columnwidth]{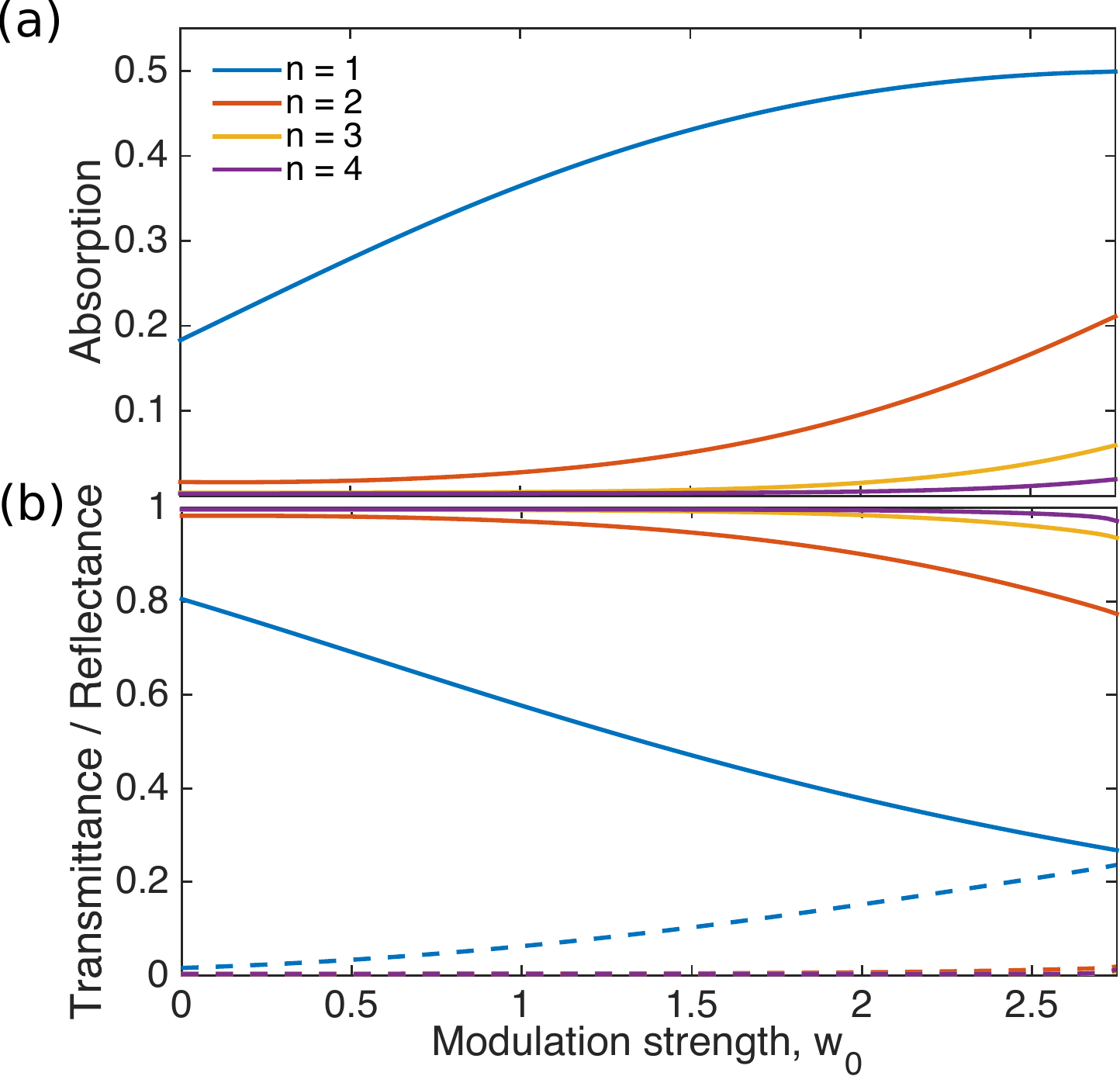}
	\caption{ Conductivity and dielectric grating. (a) Absorption at each of the resonance peaks as a function of modulation strength for the dielectric and conductivity grating. (b) Transmittance (solid lines) and reflectance (dashed lines).   }
	\label{fig7}
\end{figure}

\section{conclusions}
In this paper we have used transformation optics to study plasmons in graphene excited with the help of subwavelength dielectric gratings. We have considered gratings formed by a periodic modulation of graphene's conductivity, as well as this together with a and also a subwavelength dielectric grating of the same periodicity placed close to the graphene sheet. In both cases, the shape of the periodic profiles derive from a conformal transformation that maps a Cartesian mesh into a wavy and periodic one. However, in both cases they can be accurately approximated by conform accurately to a sinusoidal shape. We have given analytic expressions for the surface plasmon dispersion relation and the optical response at normal incidence that are exact up to the quasistatic approximation. We have shown coupling to highly confined graphene surface plasmons that provide very large field enhancements, thus increasing the low optical absorption of graphene. For the case of periodic modulation of the conductivity we have discussed the optimal conditions for coupling into the surface plasmons. Absorption as high as 35\% can be achieved with a single sheet of graphene on a substrate of permittivity $\epsilon=3$, and at the same time transmission can be cut down to less than 10\%. On the other hand, for the conductivity and dielectric grating we have found that there is no optimal modulation strength to yield a maximum coupling efficiency. Finally, we have shown that accompanying a dielectric grating with a conductivity grating of the same period is a much more efficient route to couple into the plasmons than by having only the dielectric grating.

\section*{acknowledgements}
The authors wish to thank Dr. Yu Luo for fruitful discussions. 
This work was supported by the Leverhulme Trust, the EPSRC (grant number EP/L024926/1), the Gordon and Betty Moore Foundation and the National Natural Science Foundation of China (grant number 11104200). 

\bibliography{Graphene_plasmonics,TO_Metamaterials,books}

\begin{thebibliography}{43}%
\makeatletter
\providecommand \@ifxundefined [1]{%
 \@ifx{#1\undefined}
}%
\providecommand \@ifnum [1]{%
 \ifnum #1\expandafter \@firstoftwo
 \else \expandafter \@secondoftwo
 \fi
}%
\providecommand \@ifx [1]{%
 \ifx #1\expandafter \@firstoftwo
 \else \expandafter \@secondoftwo
 \fi
}%
\providecommand \natexlab [1]{#1}%
\providecommand \enquote  [1]{``#1''}%
\providecommand \bibnamefont  [1]{#1}%
\providecommand \bibfnamefont [1]{#1}%
\providecommand \citenamefont [1]{#1}%
\providecommand \href@noop [0]{\@secondoftwo}%
\providecommand \href [0]{\begingroup \@sanitize@url \@href}%
\providecommand \@href[1]{\@@startlink{#1}\@@href}%
\providecommand \@@href[1]{\endgroup#1\@@endlink}%
\providecommand \@sanitize@url [0]{\catcode `\\12\catcode `\$12\catcode
  `\&12\catcode `\#12\catcode `\^12\catcode `\_12\catcode `\%12\relax}%
\providecommand \@@startlink[1]{}%
\providecommand \@@endlink[0]{}%
\providecommand \url  [0]{\begingroup\@sanitize@url \@url }%
\providecommand \@url [1]{\endgroup\@href {#1}{\urlprefix }}%
\providecommand \urlprefix  [0]{URL }%
\providecommand \Eprint [0]{\href }%
\providecommand \doibase [0]{http://dx.doi.org/}%
\providecommand \selectlanguage [0]{\@gobble}%
\providecommand \bibinfo  [0]{\@secondoftwo}%
\providecommand \bibfield  [0]{\@secondoftwo}%
\providecommand \translation [1]{[#1]}%
\providecommand \BibitemOpen [0]{}%
\providecommand \bibitemStop [0]{}%
\providecommand \bibitemNoStop [0]{.\EOS\space}%
\providecommand \EOS [0]{\spacefactor3000\relax}%
\providecommand \BibitemShut  [1]{\csname bibitem#1\endcsname}%
\let\auto@bib@innerbib\@empty
\bibitem [{\citenamefont {{Castro Neto}}\ \emph {et~al.}(2009)\citenamefont
  {{Castro Neto}}, \citenamefont {Peres}, \citenamefont {Novoselov},\ and\
  \citenamefont {Geim}}]{CastroNeto2009}%
  \BibitemOpen
  \bibfield  {author} {\bibinfo {author} {\bibfnamefont {A.~H.}\ \bibnamefont
  {{Castro Neto}}}, \bibinfo {author} {\bibfnamefont {N.~M.~R.}\ \bibnamefont
  {Peres}}, \bibinfo {author} {\bibfnamefont {K.~S.}\ \bibnamefont
  {Novoselov}}, \ and\ \bibinfo {author} {\bibfnamefont {A.~K.}\ \bibnamefont
  {Geim}},\ }\href {\doibase 10.1103/RevModPhys.81.109} {\bibfield  {journal}
  {\bibinfo  {journal} {Reviews of Modern Physics}\ }\textbf {\bibinfo {volume}
  {81}},\ \bibinfo {pages} {109} (\bibinfo {year} {2009})},\ \Eprint
  {http://arxiv.org/abs/0709.1163} {arXiv:0709.1163} \BibitemShut {NoStop}%
\bibitem [{\citenamefont {Bonaccorso}\ \emph {et~al.}(2010)\citenamefont
  {Bonaccorso}, \citenamefont {Sun}, \citenamefont {Hasan},\ and\ \citenamefont
  {Ferrari}}]{Bonaccorso2010}%
  \BibitemOpen
  \bibfield  {author} {\bibinfo {author} {\bibfnamefont {F.}~\bibnamefont
  {Bonaccorso}}, \bibinfo {author} {\bibfnamefont {Z.}~\bibnamefont {Sun}},
  \bibinfo {author} {\bibfnamefont {T.}~\bibnamefont {Hasan}}, \ and\ \bibinfo
  {author} {\bibfnamefont {A.~C.}\ \bibnamefont {Ferrari}},\ }\href
  {http://arxiv.org/abs/1006.4854} {\bibfield  {journal} {\bibinfo  {journal}
  {Nature Photonics}\ }\textbf {\bibinfo {volume} {4}},\ \bibinfo {pages} {611}
  (\bibinfo {year} {2010})}\BibitemShut {NoStop}%
\bibitem [{\citenamefont {Liu}\ \emph {et~al.}(2011)\citenamefont {Liu},
  \citenamefont {Yin}, \citenamefont {Ulin-Avila}, \citenamefont {Geng},
  \citenamefont {Zentgraf}, \citenamefont {Ju}, \citenamefont {Wang},\ and\
  \citenamefont {Zhang}}]{Liu2011}%
  \BibitemOpen
  \bibfield  {author} {\bibinfo {author} {\bibfnamefont {M.}~\bibnamefont
  {Liu}}, \bibinfo {author} {\bibfnamefont {X.}~\bibnamefont {Yin}}, \bibinfo
  {author} {\bibfnamefont {E.}~\bibnamefont {Ulin-Avila}}, \bibinfo {author}
  {\bibfnamefont {B.}~\bibnamefont {Geng}}, \bibinfo {author} {\bibfnamefont
  {T.}~\bibnamefont {Zentgraf}}, \bibinfo {author} {\bibfnamefont
  {L.}~\bibnamefont {Ju}}, \bibinfo {author} {\bibfnamefont {F.}~\bibnamefont
  {Wang}}, \ and\ \bibinfo {author} {\bibfnamefont {X.}~\bibnamefont {Zhang}},\
  }\href {\doibase 10.1038/nature10067} {\bibfield  {journal} {\bibinfo
  {journal} {Nature}\ }\textbf {\bibinfo {volume} {474}},\ \bibinfo {pages}
  {64} (\bibinfo {year} {2011})}\BibitemShut {NoStop}%
\bibitem [{\citenamefont {Vakil}\ and\ \citenamefont
  {Engheta}(2011)}]{Vakil2011}%
  \BibitemOpen
  \bibfield  {author} {\bibinfo {author} {\bibfnamefont {A.}~\bibnamefont
  {Vakil}}\ and\ \bibinfo {author} {\bibfnamefont {N.}~\bibnamefont
  {Engheta}},\ }\href {\doibase 10.1126/science.1202691} {\bibfield  {journal}
  {\bibinfo  {journal} {Science (New York, N.Y.)}\ }\textbf {\bibinfo {volume}
  {332}},\ \bibinfo {pages} {1291} (\bibinfo {year} {2011})}\BibitemShut
  {NoStop}%
\bibitem [{\citenamefont {Grigorenko}\ \emph {et~al.}(2012)\citenamefont
  {Grigorenko}, \citenamefont {Polini},\ and\ \citenamefont
  {Novoselov}}]{Grigorenko2012}%
  \BibitemOpen
  \bibfield  {author} {\bibinfo {author} {\bibfnamefont {a.~N.}\ \bibnamefont
  {Grigorenko}}, \bibinfo {author} {\bibfnamefont {M.}~\bibnamefont {Polini}},
  \ and\ \bibinfo {author} {\bibfnamefont {K.~S.}\ \bibnamefont {Novoselov}},\
  }\href {\doibase 10.1038/nphoton.2012.262} {\bibfield  {journal} {\bibinfo
  {journal} {Nature Photonics}\ }\textbf {\bibinfo {volume} {6}},\ \bibinfo
  {pages} {749} (\bibinfo {year} {2012})}\BibitemShut {NoStop}%
\bibitem [{\citenamefont {{Garc\'{\i}a de Abajo}}(2014)}]{GarciadeAbajo2014}%
  \BibitemOpen
  \bibfield  {author} {\bibinfo {author} {\bibfnamefont {F.~J.}\ \bibnamefont
  {{Garc\'{\i}a de Abajo}}},\ }\href@noop {} {\bibfield  {journal} {\bibinfo
  {journal} {ACS Photonics}\ }\textbf {\bibinfo {volume} {1}},\ \bibinfo
  {pages} {135} (\bibinfo {year} {2014})}\BibitemShut {NoStop}%
\bibitem [{\citenamefont {Low}\ and\ \citenamefont {Avouris}(2014)}]{Low2014}%
  \BibitemOpen
  \bibfield  {author} {\bibinfo {author} {\bibfnamefont {T.}~\bibnamefont
  {Low}}\ and\ \bibinfo {author} {\bibfnamefont {P.}~\bibnamefont {Avouris}},\
  }\href {\doibase 10.1021/nn406627u} {\bibfield  {journal} {\bibinfo
  {journal} {ACS Nano}\ }\textbf {\bibinfo {volume} {8}},\ \bibinfo {pages}
  {1086} (\bibinfo {year} {2014})},\ \Eprint {http://arxiv.org/abs/1403.2799}
  {arXiv:1403.2799} \BibitemShut {NoStop}%
\bibitem [{\citenamefont {Shung}(1986)}]{Shung1986}%
  \BibitemOpen
  \bibfield  {author} {\bibinfo {author} {\bibfnamefont {K.~W.~K.}\
  \bibnamefont {Shung}},\ }\href@noop {} {\bibfield  {journal} {\bibinfo
  {journal} {Physical Review B}\ }\textbf {\bibinfo {volume} {34}},\ \bibinfo
  {pages} {979} (\bibinfo {year} {1986})}\BibitemShut {NoStop}%
\bibitem [{\citenamefont {Vafek}(2006)}]{Vafek2006}%
  \BibitemOpen
  \bibfield  {author} {\bibinfo {author} {\bibfnamefont {O.}~\bibnamefont
  {Vafek}},\ }\href {\doibase 10.1103/PhysRevLett.97.266406} {\bibfield
  {journal} {\bibinfo  {journal} {Physical Review Letters}\ }\textbf {\bibinfo
  {volume} {97}},\ \bibinfo {pages} {266406} (\bibinfo {year}
  {2006})}\BibitemShut {NoStop}%
\bibitem [{\citenamefont {Hanson}(2008)}]{Hanson2008}%
  \BibitemOpen
  \bibfield  {author} {\bibinfo {author} {\bibfnamefont {G.~W.}\ \bibnamefont
  {Hanson}},\ }\href {\doibase 10.1063/1.2891452} {\bibfield  {journal}
  {\bibinfo  {journal} {Journal of Applied Physics}\ }\textbf {\bibinfo
  {volume} {103}},\ \bibinfo {pages} {064302} (\bibinfo {year}
  {2008})}\BibitemShut {NoStop}%
\bibitem [{\citenamefont {Jablan}\ \emph {et~al.}(2009)\citenamefont {Jablan},
  \citenamefont {Buljan},\ and\ \citenamefont {Solja\v{c}i\'{c}}}]{Jablan2009}%
  \BibitemOpen
  \bibfield  {author} {\bibinfo {author} {\bibfnamefont {M.}~\bibnamefont
  {Jablan}}, \bibinfo {author} {\bibfnamefont {H.}~\bibnamefont {Buljan}}, \
  and\ \bibinfo {author} {\bibfnamefont {M.}~\bibnamefont {Solja\v{c}i\'{c}}},\
  }\href {\doibase 10.1103/PhysRevB.80.245435} {\bibfield  {journal} {\bibinfo
  {journal} {Physical Review B}\ }\textbf {\bibinfo {volume} {80}},\ \bibinfo
  {pages} {245435} (\bibinfo {year} {2009})}\BibitemShut {NoStop}%
\bibitem [{\citenamefont {Dubinov}\ \emph {et~al.}(2011)\citenamefont
  {Dubinov}, \citenamefont {Aleshkin}, \citenamefont {Mitin}, \citenamefont
  {Otsuji},\ and\ \citenamefont {Ryzhii}}]{Dubinov2011}%
  \BibitemOpen
  \bibfield  {author} {\bibinfo {author} {\bibfnamefont {A.~A.}\ \bibnamefont
  {Dubinov}}, \bibinfo {author} {\bibfnamefont {V.~Y.}\ \bibnamefont
  {Aleshkin}}, \bibinfo {author} {\bibfnamefont {V.}~\bibnamefont {Mitin}},
  \bibinfo {author} {\bibfnamefont {T.}~\bibnamefont {Otsuji}}, \ and\ \bibinfo
  {author} {\bibfnamefont {V.}~\bibnamefont {Ryzhii}},\ }\href {\doibase
  10.1088/0953-8984/23/14/145302} {\bibfield  {journal} {\bibinfo  {journal}
  {Journal of Phys.: Condensed Matter}\ }\textbf {\bibinfo {volume} {23}},\
  \bibinfo {pages} {145302} (\bibinfo {year} {2011})}\BibitemShut {NoStop}%
\bibitem [{\citenamefont {Koppens}\ \emph {et~al.}(2011)\citenamefont
  {Koppens}, \citenamefont {Chang},\ and\ \citenamefont {{Garc\'{\i}a de
  Abajo}}}]{Koppens2011}%
  \BibitemOpen
  \bibfield  {author} {\bibinfo {author} {\bibfnamefont {F.~H.~L.}\
  \bibnamefont {Koppens}}, \bibinfo {author} {\bibfnamefont {D.~E.}\
  \bibnamefont {Chang}}, \ and\ \bibinfo {author} {\bibfnamefont {F.~J.}\
  \bibnamefont {{Garc\'{\i}a de Abajo}}},\ }\href {\doibase 10.1021/nl201771h}
  {\bibfield  {journal} {\bibinfo  {journal} {Nano letters}\ }\textbf {\bibinfo
  {volume} {11}},\ \bibinfo {pages} {3370} (\bibinfo {year}
  {2011})}\BibitemShut {NoStop}%
\bibitem [{\citenamefont {Nikitin}\ \emph {et~al.}(2011)\citenamefont
  {Nikitin}, \citenamefont {Guinea}, \citenamefont {Garc\'{\i}a-Vidal},\ and\
  \citenamefont {Mart\'{\i}n-Moreno}}]{Nikitin2011}%
  \BibitemOpen
  \bibfield  {author} {\bibinfo {author} {\bibfnamefont {A.~Y.}\ \bibnamefont
  {Nikitin}}, \bibinfo {author} {\bibfnamefont {F.}~\bibnamefont {Guinea}},
  \bibinfo {author} {\bibfnamefont {F.~J.}\ \bibnamefont {Garc\'{\i}a-Vidal}},
  \ and\ \bibinfo {author} {\bibfnamefont {L.}~\bibnamefont
  {Mart\'{\i}n-Moreno}},\ }\href {\doibase 10.1103/PhysRevB.84.195446}
  {\bibfield  {journal} {\bibinfo  {journal} {Physical Review B}\ }\textbf
  {\bibinfo {volume} {84}},\ \bibinfo {pages} {195446} (\bibinfo {year}
  {2011})}\BibitemShut {NoStop}%
\bibitem [{\citenamefont {Fei}\ \emph {et~al.}(2011)\citenamefont {Fei},
  \citenamefont {Andreev}, \citenamefont {Bao}, \citenamefont {Zhang},
  \citenamefont {{S McLeod}}, \citenamefont {Wang}, \citenamefont {Stewart},
  \citenamefont {Zhao}, \citenamefont {Dominguez}, \citenamefont {Thiemens},
  \citenamefont {Fogler}, \citenamefont {Tauber}, \citenamefont {Castro-Neto},
  \citenamefont {Lau}, \citenamefont {Keilmann},\ and\ \citenamefont
  {Basov}}]{Fei2011}%
  \BibitemOpen
  \bibfield  {author} {\bibinfo {author} {\bibfnamefont {Z.}~\bibnamefont
  {Fei}}, \bibinfo {author} {\bibfnamefont {G.~O.}\ \bibnamefont {Andreev}},
  \bibinfo {author} {\bibfnamefont {W.}~\bibnamefont {Bao}}, \bibinfo {author}
  {\bibfnamefont {L.~M.}\ \bibnamefont {Zhang}}, \bibinfo {author}
  {\bibfnamefont {A.}~\bibnamefont {{S McLeod}}}, \bibinfo {author}
  {\bibfnamefont {C.}~\bibnamefont {Wang}}, \bibinfo {author} {\bibfnamefont
  {M.~K.}\ \bibnamefont {Stewart}}, \bibinfo {author} {\bibfnamefont
  {Z.}~\bibnamefont {Zhao}}, \bibinfo {author} {\bibfnamefont {G.}~\bibnamefont
  {Dominguez}}, \bibinfo {author} {\bibfnamefont {M.}~\bibnamefont {Thiemens}},
  \bibinfo {author} {\bibfnamefont {M.~M.}\ \bibnamefont {Fogler}}, \bibinfo
  {author} {\bibfnamefont {M.~J.}\ \bibnamefont {Tauber}}, \bibinfo {author}
  {\bibfnamefont {A.~H.}\ \bibnamefont {Castro-Neto}}, \bibinfo {author}
  {\bibfnamefont {C.~N.}\ \bibnamefont {Lau}}, \bibinfo {author} {\bibfnamefont
  {F.}~\bibnamefont {Keilmann}}, \ and\ \bibinfo {author} {\bibfnamefont
  {D.~N.}\ \bibnamefont {Basov}},\ }\href {\doibase 10.1021/nl202362d}
  {\bibfield  {journal} {\bibinfo  {journal} {Nano letters}\ }\textbf {\bibinfo
  {volume} {11}},\ \bibinfo {pages} {4701} (\bibinfo {year}
  {2011})}\BibitemShut {NoStop}%
\bibitem [{\citenamefont {Fei}\ \emph {et~al.}(2012)\citenamefont {Fei},
  \citenamefont {Rodin}, \citenamefont {Andreev}, \citenamefont {Bao},
  \citenamefont {McLeod}, \citenamefont {Wagner}, \citenamefont {Zhang},
  \citenamefont {Zhao}, \citenamefont {Thiemens}, \citenamefont {Dominguez},
  \citenamefont {Fogler}, \citenamefont {{Castro Neto}}, \citenamefont {Lau},
  \citenamefont {Keilmann},\ and\ \citenamefont {Basov}}]{Fei2012}%
  \BibitemOpen
  \bibfield  {author} {\bibinfo {author} {\bibfnamefont {Z.}~\bibnamefont
  {Fei}}, \bibinfo {author} {\bibfnamefont {a.~S.}\ \bibnamefont {Rodin}},
  \bibinfo {author} {\bibfnamefont {G.~O.}\ \bibnamefont {Andreev}}, \bibinfo
  {author} {\bibfnamefont {W.}~\bibnamefont {Bao}}, \bibinfo {author}
  {\bibfnamefont {a.~S.}\ \bibnamefont {McLeod}}, \bibinfo {author}
  {\bibfnamefont {M.}~\bibnamefont {Wagner}}, \bibinfo {author} {\bibfnamefont
  {L.~M.}\ \bibnamefont {Zhang}}, \bibinfo {author} {\bibfnamefont
  {Z.}~\bibnamefont {Zhao}}, \bibinfo {author} {\bibfnamefont {M.}~\bibnamefont
  {Thiemens}}, \bibinfo {author} {\bibfnamefont {G.}~\bibnamefont {Dominguez}},
  \bibinfo {author} {\bibfnamefont {M.~M.}\ \bibnamefont {Fogler}}, \bibinfo
  {author} {\bibfnamefont {a.~H.}\ \bibnamefont {{Castro Neto}}}, \bibinfo
  {author} {\bibfnamefont {C.~N.}\ \bibnamefont {Lau}}, \bibinfo {author}
  {\bibfnamefont {F.}~\bibnamefont {Keilmann}}, \ and\ \bibinfo {author}
  {\bibfnamefont {D.~N.}\ \bibnamefont {Basov}},\ }\href {\doibase
  10.1038/nature11253} {\bibfield  {journal} {\bibinfo  {journal} {Nature}\
  }\textbf {\bibinfo {volume} {487}},\ \bibinfo {pages} {82} (\bibinfo {year}
  {2012})}\BibitemShut {NoStop}%
\bibitem [{\citenamefont {Chen}\ \emph {et~al.}(2012)\citenamefont {Chen},
  \citenamefont {Badioli}, \citenamefont {Alonso-Gonz\'{a}lez}, \citenamefont
  {Thongrattanasiri}, \citenamefont {Huth}, \citenamefont {Osmond},
  \citenamefont {Spasenovi\'{c}}, \citenamefont {Centeno}, \citenamefont
  {Pesquera}, \citenamefont {Godignon}, \citenamefont {Elorza}, \citenamefont
  {Camara}, \citenamefont {{Garc\'{\i}a de Abajo}}, \citenamefont
  {Hillenbrand},\ and\ \citenamefont {Koppens}}]{Chen2012}%
  \BibitemOpen
  \bibfield  {author} {\bibinfo {author} {\bibfnamefont {J.}~\bibnamefont
  {Chen}}, \bibinfo {author} {\bibfnamefont {M.}~\bibnamefont {Badioli}},
  \bibinfo {author} {\bibfnamefont {P.}~\bibnamefont {Alonso-Gonz\'{a}lez}},
  \bibinfo {author} {\bibfnamefont {S.}~\bibnamefont {Thongrattanasiri}},
  \bibinfo {author} {\bibfnamefont {F.}~\bibnamefont {Huth}}, \bibinfo {author}
  {\bibfnamefont {J.}~\bibnamefont {Osmond}}, \bibinfo {author} {\bibfnamefont
  {M.}~\bibnamefont {Spasenovi\'{c}}}, \bibinfo {author} {\bibfnamefont
  {A.}~\bibnamefont {Centeno}}, \bibinfo {author} {\bibfnamefont
  {A.}~\bibnamefont {Pesquera}}, \bibinfo {author} {\bibfnamefont
  {P.}~\bibnamefont {Godignon}}, \bibinfo {author} {\bibfnamefont {A.~Z.}\
  \bibnamefont {Elorza}}, \bibinfo {author} {\bibfnamefont {N.}~\bibnamefont
  {Camara}}, \bibinfo {author} {\bibfnamefont {F.~J.}\ \bibnamefont
  {{Garc\'{\i}a de Abajo}}}, \bibinfo {author} {\bibfnamefont {R.}~\bibnamefont
  {Hillenbrand}}, \ and\ \bibinfo {author} {\bibfnamefont {F.~H.~L.}\
  \bibnamefont {Koppens}},\ }\href {\doibase 10.1038/nature11254} {\bibfield
  {journal} {\bibinfo  {journal} {Nature}\ }\textbf {\bibinfo {volume} {487}},\
  \bibinfo {pages} {77} (\bibinfo {year} {2012})}\BibitemShut {NoStop}%
\bibitem [{\citenamefont {Bludov}\ \emph {et~al.}(2012)\citenamefont {Bludov},
  \citenamefont {Peres},\ and\ \citenamefont {Vasilevskiy}}]{Bludov2012}%
  \BibitemOpen
  \bibfield  {author} {\bibinfo {author} {\bibfnamefont {Y.~V.}\ \bibnamefont
  {Bludov}}, \bibinfo {author} {\bibfnamefont {N.~M.~R.}\ \bibnamefont
  {Peres}}, \ and\ \bibinfo {author} {\bibfnamefont {M.~I.}\ \bibnamefont
  {Vasilevskiy}},\ }\href {\doibase 10.1103/PhysRevB.85.245409} {\bibfield
  {journal} {\bibinfo  {journal} {Physical Review B - Condensed Matter and
  Materials Physics}\ }\textbf {\bibinfo {volume} {85}},\ \bibinfo {pages} {1}
  (\bibinfo {year} {2012})},\ \Eprint {http://arxiv.org/abs/arXiv:1204.3900v1}
  {arXiv:arXiv:1204.3900v1} \BibitemShut {NoStop}%
\bibitem [{\citenamefont {Alonso-Gonz\'{a}lez}\ \emph
  {et~al.}(2014)\citenamefont {Alonso-Gonz\'{a}lez}, \citenamefont {Nikitin},
  \citenamefont {Golmar}, \citenamefont {Centeno}, \citenamefont {Pesquera},
  \citenamefont {V\'{e}lez}, \citenamefont {Chen}, \citenamefont {Navickaite},
  \citenamefont {Koppens}, \citenamefont {Zurutuza}, \citenamefont {Casanova},
  \citenamefont {Hueso},\ and\ \citenamefont
  {Hillenbrand}}]{Alonso-Gonzalez2014}%
  \BibitemOpen
  \bibfield  {author} {\bibinfo {author} {\bibfnamefont {P.}~\bibnamefont
  {Alonso-Gonz\'{a}lez}}, \bibinfo {author} {\bibfnamefont {a.~Y.}\
  \bibnamefont {Nikitin}}, \bibinfo {author} {\bibfnamefont {F.}~\bibnamefont
  {Golmar}}, \bibinfo {author} {\bibfnamefont {a.}~\bibnamefont {Centeno}},
  \bibinfo {author} {\bibfnamefont {a.}~\bibnamefont {Pesquera}}, \bibinfo
  {author} {\bibfnamefont {S.}~\bibnamefont {V\'{e}lez}}, \bibinfo {author}
  {\bibfnamefont {J.}~\bibnamefont {Chen}}, \bibinfo {author} {\bibfnamefont
  {G.}~\bibnamefont {Navickaite}}, \bibinfo {author} {\bibfnamefont
  {F.}~\bibnamefont {Koppens}}, \bibinfo {author} {\bibfnamefont
  {a.}~\bibnamefont {Zurutuza}}, \bibinfo {author} {\bibfnamefont
  {F.}~\bibnamefont {Casanova}}, \bibinfo {author} {\bibfnamefont {L.~E.}\
  \bibnamefont {Hueso}}, \ and\ \bibinfo {author} {\bibfnamefont
  {R.}~\bibnamefont {Hillenbrand}},\ }\href {\doibase 10.1126/science.1253202}
  {\bibfield  {journal} {\bibinfo  {journal} {Science (New York, N.Y.)}\
  }\textbf {\bibinfo {volume} {344}},\ \bibinfo {pages} {1369} (\bibinfo {year}
  {2014})}\BibitemShut {NoStop}%
\bibitem [{\citenamefont {Slipchenko}\ \emph {et~al.}(2013)\citenamefont
  {Slipchenko}, \citenamefont {Nesterov}, \citenamefont {Martin-Moreno},\ and\
  \citenamefont {Nikitin}}]{Slipchenko2013}%
  \BibitemOpen
  \bibfield  {author} {\bibinfo {author} {\bibfnamefont {T.~M.}\ \bibnamefont
  {Slipchenko}}, \bibinfo {author} {\bibfnamefont {M.~L.}\ \bibnamefont
  {Nesterov}}, \bibinfo {author} {\bibfnamefont {L.}~\bibnamefont
  {Martin-Moreno}}, \ and\ \bibinfo {author} {\bibfnamefont {a.~Y.}\
  \bibnamefont {Nikitin}},\ }\href {\doibase 10.1088/2040-8978/15/11/114008}
  {\bibfield  {journal} {\bibinfo  {journal} {Journal of Optics}\ }\textbf
  {\bibinfo {volume} {15}},\ \bibinfo {pages} {114008} (\bibinfo {year}
  {2013})},\ \Eprint {http://arxiv.org/abs/1307.0310} {arXiv:1307.0310}
  \BibitemShut {NoStop}%
\bibitem [{\citenamefont {Smirnova}\ \emph {et~al.}(2015)\citenamefont
  {Smirnova}, \citenamefont {Mousavi}, \citenamefont {Kivshar},\ and\
  \citenamefont {Khanikaev}}]{Smirnova2015}%
  \BibitemOpen
  \bibfield  {author} {\bibinfo {author} {\bibfnamefont {D.}~\bibnamefont
  {Smirnova}}, \bibinfo {author} {\bibfnamefont {H.}~\bibnamefont {Mousavi}},
  \bibinfo {author} {\bibfnamefont {Y.~S.}\ \bibnamefont {Kivshar}}, \ and\
  \bibinfo {author} {\bibfnamefont {A.~B.}\ \bibnamefont {Khanikaev}},\
  }\href@noop {} {\bibfield  {journal} {\bibinfo  {journal}
  {arxiv:15:08.02729v1}\ } (\bibinfo {year} {2015})}\BibitemShut {NoStop}%
\bibitem [{\citenamefont {Zhan}\ \emph {et~al.}(2012)\citenamefont {Zhan},
  \citenamefont {Zhao}, \citenamefont {Hu}, \citenamefont {Liu},\ and\
  \citenamefont {Zi}}]{Zhan2012}%
  \BibitemOpen
  \bibfield  {author} {\bibinfo {author} {\bibfnamefont {T.~R.}\ \bibnamefont
  {Zhan}}, \bibinfo {author} {\bibfnamefont {F.~Y.}\ \bibnamefont {Zhao}},
  \bibinfo {author} {\bibfnamefont {X.~H.}\ \bibnamefont {Hu}}, \bibinfo
  {author} {\bibfnamefont {X.~H.}\ \bibnamefont {Liu}}, \ and\ \bibinfo
  {author} {\bibfnamefont {J.}~\bibnamefont {Zi}},\ }\href {\doibase
  10.1103/PhysRevB.86.165416} {\bibfield  {journal} {\bibinfo  {journal}
  {Physical Review B - Condensed Matter and Materials Physics}\ }\textbf
  {\bibinfo {volume} {86}},\ \bibinfo {pages} {2} (\bibinfo {year}
  {2012})}\BibitemShut {NoStop}%
\bibitem [{\citenamefont {Ju}\ \emph {et~al.}(2011)\citenamefont {Ju},
  \citenamefont {Geng}, \citenamefont {Horng}, \citenamefont {Girit},
  \citenamefont {Martin}, \citenamefont {Hao}, \citenamefont {Bechtel},
  \citenamefont {Liang}, \citenamefont {Zettl}, \citenamefont {Shen},\ and\
  \citenamefont {Wang}}]{Ju2011}%
  \BibitemOpen
  \bibfield  {author} {\bibinfo {author} {\bibfnamefont {L.}~\bibnamefont
  {Ju}}, \bibinfo {author} {\bibfnamefont {B.}~\bibnamefont {Geng}}, \bibinfo
  {author} {\bibfnamefont {J.}~\bibnamefont {Horng}}, \bibinfo {author}
  {\bibfnamefont {C.}~\bibnamefont {Girit}}, \bibinfo {author} {\bibfnamefont
  {M.}~\bibnamefont {Martin}}, \bibinfo {author} {\bibfnamefont
  {Z.}~\bibnamefont {Hao}}, \bibinfo {author} {\bibfnamefont {H.~a.}\
  \bibnamefont {Bechtel}}, \bibinfo {author} {\bibfnamefont {X.}~\bibnamefont
  {Liang}}, \bibinfo {author} {\bibfnamefont {A.}~\bibnamefont {Zettl}},
  \bibinfo {author} {\bibfnamefont {Y.~R.}\ \bibnamefont {Shen}}, \ and\
  \bibinfo {author} {\bibfnamefont {F.}~\bibnamefont {Wang}},\ }\href {\doibase
  10.1038/nnano.2011.146} {\bibfield  {journal} {\bibinfo  {journal} {Nature
  nanotechnology}\ }\textbf {\bibinfo {volume} {6}},\ \bibinfo {pages} {630}
  (\bibinfo {year} {2011})}\BibitemShut {NoStop}%
\bibitem [{\citenamefont {Nikitin}\ \emph
  {et~al.}(2012{\natexlab{a}})\citenamefont {Nikitin}, \citenamefont {Guinea},
  \citenamefont {Garcia-Vidal},\ and\ \citenamefont
  {Martin-Moreno}}]{Nikitin2012b}%
  \BibitemOpen
  \bibfield  {author} {\bibinfo {author} {\bibfnamefont {a.~Y.}\ \bibnamefont
  {Nikitin}}, \bibinfo {author} {\bibfnamefont {F.}~\bibnamefont {Guinea}},
  \bibinfo {author} {\bibfnamefont {F.~J.}\ \bibnamefont {Garcia-Vidal}}, \
  and\ \bibinfo {author} {\bibfnamefont {L.}~\bibnamefont {Martin-Moreno}},\
  }\href {\doibase 10.1103/PhysRevB.85.081405} {\bibfield  {journal} {\bibinfo
  {journal} {Physical Review B - Condensed Matter and Materials Physics}\
  }\textbf {\bibinfo {volume} {85}},\ \bibinfo {pages} {1} (\bibinfo {year}
  {2012}{\natexlab{a}})},\ \Eprint {http://arxiv.org/abs/1201.0191}
  {arXiv:1201.0191 [cond-mat.mes-hall]} \BibitemShut {NoStop}%
\bibitem [{\citenamefont {Nikitin}\ \emph {et~al.}(2014)\citenamefont
  {Nikitin}, \citenamefont {Low},\ and\ \citenamefont
  {Martin-Moreno}}]{Nikitin2014}%
  \BibitemOpen
  \bibfield  {author} {\bibinfo {author} {\bibfnamefont {a.~Y.}\ \bibnamefont
  {Nikitin}}, \bibinfo {author} {\bibfnamefont {T.}~\bibnamefont {Low}}, \ and\
  \bibinfo {author} {\bibfnamefont {L.}~\bibnamefont {Martin-Moreno}},\ }\href
  {\doibase 10.1103/PhysRevB.90.041407} {\bibfield  {journal} {\bibinfo
  {journal} {Phys. Rev. B}\ }\textbf {\bibinfo {volume} {90}},\ \bibinfo
  {pages} {041407} (\bibinfo {year} {2014})},\ \Eprint
  {http://arxiv.org/abs/arXiv:1406.7335v1} {arXiv:arXiv:1406.7335v1}
  \BibitemShut {NoStop}%
\bibitem [{\citenamefont {Nikitin}\ \emph
  {et~al.}(2012{\natexlab{b}})\citenamefont {Nikitin}, \citenamefont {Guinea},\
  and\ \citenamefont {Mart\'{\i}n-Moreno}}]{Nikitin2012}%
  \BibitemOpen
  \bibfield  {author} {\bibinfo {author} {\bibfnamefont {A.~Y.}\ \bibnamefont
  {Nikitin}}, \bibinfo {author} {\bibfnamefont {F.}~\bibnamefont {Guinea}}, \
  and\ \bibinfo {author} {\bibfnamefont {L.}~\bibnamefont
  {Mart\'{\i}n-Moreno}},\ }\href {\doibase 10.1063/1.4760230} {\bibfield
  {journal} {\bibinfo  {journal} {Applied Physics Letters}\ }\textbf {\bibinfo
  {volume} {101}},\ \bibinfo {pages} {151119} (\bibinfo {year}
  {2012}{\natexlab{b}})}\BibitemShut {NoStop}%
\bibitem [{\citenamefont {Thongrattanasiri}\ \emph {et~al.}(2012)\citenamefont
  {Thongrattanasiri}, \citenamefont {Koppens},\ and\ \citenamefont
  {{Garc\'{\i}a De Abajo}}}]{Thongrattanasiri2012a}%
  \BibitemOpen
  \bibfield  {author} {\bibinfo {author} {\bibfnamefont {S.}~\bibnamefont
  {Thongrattanasiri}}, \bibinfo {author} {\bibfnamefont {F.~H.~L.}\
  \bibnamefont {Koppens}}, \ and\ \bibinfo {author} {\bibfnamefont {F.~J.}\
  \bibnamefont {{Garc\'{\i}a De Abajo}}},\ }\href {\doibase
  10.1103/PhysRevLett.108.047401} {\bibfield  {journal} {\bibinfo  {journal}
  {Physical Review Letters}\ }\textbf {\bibinfo {volume} {108}},\ \bibinfo
  {pages} {1} (\bibinfo {year} {2012})},\ \Eprint
  {http://arxiv.org/abs/1106.4460} {arXiv:1106.4460} \BibitemShut {NoStop}%
\bibitem [{\citenamefont {Stauber}\ \emph {et~al.}(2014)\citenamefont
  {Stauber}, \citenamefont {G\'{o}mez-Santos},\ and\ \citenamefont {{De
  Abajo}}}]{Stauber2014a}%
  \BibitemOpen
  \bibfield  {author} {\bibinfo {author} {\bibfnamefont {T.}~\bibnamefont
  {Stauber}}, \bibinfo {author} {\bibfnamefont {G.}~\bibnamefont
  {G\'{o}mez-Santos}}, \ and\ \bibinfo {author} {\bibfnamefont {F.~J.~G.}\
  \bibnamefont {{De Abajo}}},\ }\href {\doibase 10.1103/PhysRevLett.112.077401}
  {\bibfield  {journal} {\bibinfo  {journal} {Physical Review Letters}\
  }\textbf {\bibinfo {volume} {112}} (\bibinfo {year} {2014}),\
  10.1103/PhysRevLett.112.077401},\ \Eprint {http://arxiv.org/abs/1310.6197}
  {arXiv:1310.6197} \BibitemShut {NoStop}%
\bibitem [{\citenamefont {Ward}\ and\ \citenamefont {Pendry}(1996)}]{Ward1996}%
  \BibitemOpen
  \bibfield  {author} {\bibinfo {author} {\bibfnamefont {A.~J.}\ \bibnamefont
  {Ward}}\ and\ \bibinfo {author} {\bibfnamefont {J.~B.}\ \bibnamefont
  {Pendry}},\ }\href@noop {} {\bibfield  {journal} {\bibinfo  {journal}
  {Journal of Modern Optics}\ }\textbf {\bibinfo {volume} {43}},\ \bibinfo
  {pages} {773 } (\bibinfo {year} {1996})}\BibitemShut {NoStop}%
\bibitem [{\citenamefont {Pendry}\ \emph {et~al.}(2006)\citenamefont {Pendry},
  \citenamefont {Schurig},\ and\ \citenamefont {Smith}}]{Pendry2006}%
  \BibitemOpen
  \bibfield  {author} {\bibinfo {author} {\bibfnamefont {J.~B.}\ \bibnamefont
  {Pendry}}, \bibinfo {author} {\bibfnamefont {D.}~\bibnamefont {Schurig}}, \
  and\ \bibinfo {author} {\bibfnamefont {D.~R.}\ \bibnamefont {Smith}},\ }\href
  {\doibase 10.1126/science.1125907} {\bibfield  {journal} {\bibinfo  {journal}
  {Science (New York, N.Y.)}\ }\textbf {\bibinfo {volume} {312}},\ \bibinfo
  {pages} {1780} (\bibinfo {year} {2006})}\BibitemShut {NoStop}%
\bibitem [{\citenamefont {Pendry}\ \emph {et~al.}(2012)\citenamefont {Pendry},
  \citenamefont {Aubry}, \citenamefont {Smith},\ and\ \citenamefont
  {Maier}}]{Pendry2012}%
  \BibitemOpen
  \bibfield  {author} {\bibinfo {author} {\bibfnamefont {J.~B.}\ \bibnamefont
  {Pendry}}, \bibinfo {author} {\bibfnamefont {A.}~\bibnamefont {Aubry}},
  \bibinfo {author} {\bibfnamefont {D.~R.}\ \bibnamefont {Smith}}, \ and\
  \bibinfo {author} {\bibfnamefont {S.~A.}\ \bibnamefont {Maier}},\ }\href
  {\doibase 10.1126/science.1220600} {\bibfield  {journal} {\bibinfo  {journal}
  {Science (New York, N.Y.)}\ }\textbf {\bibinfo {volume} {337}},\ \bibinfo
  {pages} {549} (\bibinfo {year} {2012})}\BibitemShut {NoStop}%
\bibitem [{\citenamefont {Luo}\ \emph {et~al.}(2014{\natexlab{a}})\citenamefont
  {Luo}, \citenamefont {Zhao}, \citenamefont {Fern\'{a}ndez-Dom\'{\i}nguez},
  \citenamefont {Maier},\ and\ \citenamefont {Pendry}}]{Luo2014}%
  \BibitemOpen
  \bibfield  {author} {\bibinfo {author} {\bibfnamefont {Y.}~\bibnamefont
  {Luo}}, \bibinfo {author} {\bibfnamefont {R.}~\bibnamefont {Zhao}}, \bibinfo
  {author} {\bibfnamefont {A.~I.}\ \bibnamefont
  {Fern\'{a}ndez-Dom\'{\i}nguez}}, \bibinfo {author} {\bibfnamefont {S.~A.}\
  \bibnamefont {Maier}}, \ and\ \bibinfo {author} {\bibfnamefont {J.~B.}\
  \bibnamefont {Pendry}},\ }\href
  {http://download.springer.com/static/pdf/142/art\%3A10.1007\%2Fs11432-013-5031-2.pdf?auth66=1415624115\_2969f5cf18371bbcf8221c160c67f4d9\&ext=.pdf}
  {\bibfield  {journal} {\bibinfo  {journal} {Science China Information
  Sciences}\ }\textbf {\bibinfo {volume} {56}},\ \bibinfo {pages} {120401:1}
  (\bibinfo {year} {2014}{\natexlab{a}})}\BibitemShut {NoStop}%
\bibitem [{\citenamefont {Aubry}\ \emph {et~al.}(2010)\citenamefont {Aubry},
  \citenamefont {Lei}, \citenamefont {Maier},\ and\ \citenamefont
  {Pendry}}]{Aubry2010c}%
  \BibitemOpen
  \bibfield  {author} {\bibinfo {author} {\bibfnamefont {A.}~\bibnamefont
  {Aubry}}, \bibinfo {author} {\bibfnamefont {D.~Y.}\ \bibnamefont {Lei}},
  \bibinfo {author} {\bibfnamefont {S.~A.}\ \bibnamefont {Maier}}, \ and\
  \bibinfo {author} {\bibfnamefont {J.~B.}\ \bibnamefont {Pendry}},\ }\href
  {\doibase 10.1103/PhysRevLett.105.233901} {\bibfield  {journal} {\bibinfo
  {journal} {Physical Review Letters}\ }\textbf {\bibinfo {volume} {105}},\
  \bibinfo {pages} {233901} (\bibinfo {year} {2010})}\BibitemShut {NoStop}%
\bibitem [{\citenamefont {Huidobro}\ \emph {et~al.}(2010)\citenamefont
  {Huidobro}, \citenamefont {Nesterov}, \citenamefont {Mart\'{\i}n-Moreno},\
  and\ \citenamefont {Garc\'{\i}a-Vidal}}]{Huidobro2010}%
  \BibitemOpen
  \bibfield  {author} {\bibinfo {author} {\bibfnamefont {P.~A.}\ \bibnamefont
  {Huidobro}}, \bibinfo {author} {\bibfnamefont {M.~L.}\ \bibnamefont
  {Nesterov}}, \bibinfo {author} {\bibfnamefont {L.}~\bibnamefont
  {Mart\'{\i}n-Moreno}}, \ and\ \bibinfo {author} {\bibfnamefont {F.~J.}\
  \bibnamefont {Garc\'{\i}a-Vidal}},\ }\href {\doibase 10.1021/nl100800c}
  {\bibfield  {journal} {\bibinfo  {journal} {Nano Letters}\ }\textbf {\bibinfo
  {volume} {10}},\ \bibinfo {pages} {1985} (\bibinfo {year}
  {2010})}\BibitemShut {NoStop}%
\bibitem [{\citenamefont {Liu}\ \emph {et~al.}(2010)\citenamefont {Liu},
  \citenamefont {Zentgraf}, \citenamefont {Bartal},\ and\ \citenamefont
  {Zhang}}]{Liu2010}%
  \BibitemOpen
  \bibfield  {author} {\bibinfo {author} {\bibfnamefont {Y.}~\bibnamefont
  {Liu}}, \bibinfo {author} {\bibfnamefont {T.}~\bibnamefont {Zentgraf}},
  \bibinfo {author} {\bibfnamefont {G.}~\bibnamefont {Bartal}}, \ and\ \bibinfo
  {author} {\bibfnamefont {X.}~\bibnamefont {Zhang}},\ }\href {\doibase
  10.1021/nl1008019} {\bibfield  {journal} {\bibinfo  {journal} {Nano letters}\
  }\textbf {\bibinfo {volume} {10}},\ \bibinfo {pages} {1991} (\bibinfo {year}
  {2010})}\BibitemShut {NoStop}%
\bibitem [{\citenamefont {Pendry}\ \emph {et~al.}(2013)\citenamefont {Pendry},
  \citenamefont {Fern\'{a}ndez-Dom\'{\i}nguez}, \citenamefont {Luo},\ and\
  \citenamefont {Zhao}}]{Pendry2013}%
  \BibitemOpen
  \bibfield  {author} {\bibinfo {author} {\bibfnamefont {J.~B.}\ \bibnamefont
  {Pendry}}, \bibinfo {author} {\bibfnamefont {A.~I.}\ \bibnamefont
  {Fern\'{a}ndez-Dom\'{\i}nguez}}, \bibinfo {author} {\bibfnamefont
  {Y.}~\bibnamefont {Luo}}, \ and\ \bibinfo {author} {\bibfnamefont
  {R.}~\bibnamefont {Zhao}},\ }\href {\doibase 10.1038/nphys2667} {\bibfield
  {journal} {\bibinfo  {journal} {Nature Physics}\ }\textbf {\bibinfo {volume}
  {9}},\ \bibinfo {pages} {518} (\bibinfo {year} {2013})}\BibitemShut {NoStop}%
\bibitem [{\citenamefont {Kraft}\ \emph {et~al.}(2014)\citenamefont {Kraft},
  \citenamefont {Pendry}, \citenamefont {Maier},\ and\ \citenamefont
  {Luo}}]{Kraft2014}%
  \BibitemOpen
  \bibfield  {author} {\bibinfo {author} {\bibfnamefont {M.}~\bibnamefont
  {Kraft}}, \bibinfo {author} {\bibfnamefont {J.~B.}\ \bibnamefont {Pendry}},
  \bibinfo {author} {\bibfnamefont {S.~A.}\ \bibnamefont {Maier}}, \ and\
  \bibinfo {author} {\bibfnamefont {Y.}~\bibnamefont {Luo}},\ }\href {\doibase
  10.1103/PhysRevB.89.245125} {\bibfield  {journal} {\bibinfo  {journal}
  {Physical Review B}\ }\textbf {\bibinfo {volume} {89}},\ \bibinfo {pages}
  {245125} (\bibinfo {year} {2014})}\BibitemShut {NoStop}%
\bibitem [{\citenamefont {Luo}\ \emph {et~al.}(2014{\natexlab{b}})\citenamefont
  {Luo}, \citenamefont {Zhao},\ and\ \citenamefont {Pendry}}]{Luo2014a}%
  \BibitemOpen
  \bibfield  {author} {\bibinfo {author} {\bibfnamefont {Y.}~\bibnamefont
  {Luo}}, \bibinfo {author} {\bibfnamefont {R.}~\bibnamefont {Zhao}}, \ and\
  \bibinfo {author} {\bibfnamefont {J.~B.}\ \bibnamefont {Pendry}},\ }\href
  {\doibase 10.1073/pnas.1420551111} {\bibfield  {journal} {\bibinfo  {journal}
  {Proceedings of the National Academy of Sciences}\ }\textbf {\bibinfo
  {volume} {111}},\ \bibinfo {pages} {18422} (\bibinfo {year}
  {2014}{\natexlab{b}})}\BibitemShut {NoStop}%
\bibitem [{\citenamefont {Kraft}\ \emph {et~al.}(2015)\citenamefont {Kraft},
  \citenamefont {Luo}, \citenamefont {Maier},\ and\ \citenamefont
  {Pendry}}]{Kraft2015}%
  \BibitemOpen
  \bibfield  {author} {\bibinfo {author} {\bibfnamefont {M.}~\bibnamefont
  {Kraft}}, \bibinfo {author} {\bibfnamefont {Y.}~\bibnamefont {Luo}}, \bibinfo
  {author} {\bibfnamefont {S.}~\bibnamefont {Maier}}, \ and\ \bibinfo {author}
  {\bibfnamefont {J.}~\bibnamefont {Pendry}},\ }\href {\doibase
  10.1103/PhysRevX.5.031029} {\bibfield  {journal} {\bibinfo  {journal}
  {Physical Review X}\ }\textbf {\bibinfo {volume} {5}},\ \bibinfo {pages}
  {031029} (\bibinfo {year} {2015})}\BibitemShut {NoStop}%
\bibitem [{\citenamefont {Pendry}\ and\ \citenamefont
  {Ramakrishna}(2002)}]{Pendry2002}%
  \BibitemOpen
  \bibfield  {author} {\bibinfo {author} {\bibfnamefont {J.~B.}\ \bibnamefont
  {Pendry}}\ and\ \bibinfo {author} {\bibfnamefont {S.~A.}\ \bibnamefont
  {Ramakrishna}},\ }\href {\doibase 10.1088/0953-8984/14/36/306} {\bibfield
  {journal} {\bibinfo  {journal} {Journal of Physics: Condensed Matter}\
  }\textbf {\bibinfo {volume} {14}},\ \bibinfo {pages} {8463} (\bibinfo {year}
  {2002})}\BibitemShut {NoStop}%
\bibitem [{\citenamefont {Schinzinger}\ and\ \citenamefont
  {Laura}(2003)}]{ConfMap}%
  \BibitemOpen
  \bibfield  {author} {\bibinfo {author} {\bibfnamefont {R.}~\bibnamefont
  {Schinzinger}}\ and\ \bibinfo {author} {\bibfnamefont {P.~A.~A.}\
  \bibnamefont {Laura}},\ }\href@noop {} {\emph {\bibinfo {title} {Conformal
  Mapping - Methods and Applications}}}\ (\bibinfo  {publisher} {Dover
  Publications, Inc},\ \bibinfo {address} {Minealo, New York},\ \bibinfo {year}
  {2003})\BibitemShut {NoStop}%
\bibitem [{\citenamefont {Wunsch}\ \emph {et~al.}(2006)\citenamefont {Wunsch},
  \citenamefont {Stauber}, \citenamefont {Sols},\ and\ \citenamefont
  {Guinea}}]{wunsch2006}%
  \BibitemOpen
  \bibfield  {author} {\bibinfo {author} {\bibfnamefont {B.}~\bibnamefont
  {Wunsch}}, \bibinfo {author} {\bibfnamefont {T.}~\bibnamefont {Stauber}},
  \bibinfo {author} {\bibfnamefont {F.}~\bibnamefont {Sols}}, \ and\ \bibinfo
  {author} {\bibfnamefont {F.}~\bibnamefont {Guinea}},\ }\href {\doibase
  10.1088/1367-2630/8/12/318} {\bibfield  {journal} {\bibinfo  {journal} {New
  Journal of Physics}\ }\textbf {\bibinfo {volume} {8}},\ \bibinfo {pages}
  {318} (\bibinfo {year} {2006})}\BibitemShut {NoStop}%
\bibitem [{\citenamefont {Leonhardt}\ and\ \citenamefont
  {Philbin}(2009)}]{Leonhardt2008a}%
  \BibitemOpen
  \bibfield  {author} {\bibinfo {author} {\bibfnamefont {U.}~\bibnamefont
  {Leonhardt}}\ and\ \bibinfo {author} {\bibfnamefont {T.~G.}\ \bibnamefont
  {Philbin}},\ }\href {\doibase 10.1016/S0079-6638(08)00202-3} {\bibfield
  {journal} {\bibinfo  {journal} {Progress in Optics}\ }\textbf {\bibinfo
  {volume} {53}},\ \bibinfo {pages} {69} (\bibinfo {year} {2009})},\ \Eprint
  {http://arxiv.org/abs/arXiv:0805.4778v2} {arXiv:arXiv:0805.4778v2}
  \BibitemShut {NoStop}%
\end{thebibliography}%

\end{document}